\documentclass[a4paper,12pt]{article}
\usepackage{jheppub} 
\usepackage{lineno}
\usepackage{physics}
\usepackage[english,activeacute]{babel}
\usepackage{mathtools}
\usepackage{xcolor}
\usepackage[scr=boondox,  
            cal=esstix]   
           {mathalpha}
\usepackage{tikz}
\usepackage{tikz-cd}
\usetikzlibrary{calc,patterns,angles,quotes}
\usetikzlibrary{arrows,shapes}
\usepackage{times}
\usepackage{subfig}
\usepackage[all]{xy}
\usepackage{caption}
\usepackage{soul} 

\newcommand\restr[2]{{
		\left.\kern-\nulldelimiterspace 
		#1 
		\vphantom{\big|} 
		\right|_{#2} 
}}

\title{\boldmath Pleba\'nski-Demia\'nski solutions in bigravity and Kerr-Schild double copy relations using an effective metric}

 \author[1]{Hugo Garc\'{\i}a-Compe\'an\note{hugo.compean@cinvestav.mx}}
 \author[2]{and C\'esar I. Ramos\note{cesar.ramos@cinvestav.mx}}
 \affiliation{Departamento de F\'{\i}sica,\\
 Centro de Investigaci\'on y de Estudios Avanzados del Instituto Polit\'ecnico Nacional,\\
 P.O. box 14-740, C.P. 07000, Ciudad de M\'exico, Mexico}

\abstract{In this work a formalism for proportional generalized double Kerr-Schild ansatz in bigravity is considered, where both metrics are coupled to matter. We study time-dependent and stationary solutions in the framework of the Kerr-Schild classical double copy and obtain the classical Kerr-Schild for the double, single and zeroth copy equations. For the time-dependent case, we use AdS waves solutions in bigravity previously studied in the literature. For the stationary case, we discuss a kind of Pleba\'nski-Demia\'nski solutions in bigravity which permit different masses, NUT parameters, electric and magnetic charges, while the kinematical parameters are the same, and the cosmological constants related. These solution is presented in Pleba\'nski coordinates, and it is noticed that in these coordinates the description simplifies the classical double copy equations allowing a clearer interpretation in terms of the defined fields. We present and interpret some cases for these solutions for the separate matter sector and using the effective metric.}

\makeatletter
\gdef\@fpheader{}
\makeatother

\begin{document}
\maketitle
\flushbottom
\section{Introduction}
The road to quantum gravity has many faces (for a review, see \cite{Armas:2021yut}). Among a number of them, one of the most important explored possibilities is the one of the fact that Einstein's theory of General Relativity (GR) would be modified as a macroscopic effect due to quantum corrections due to the microscopic quantum dynamics of the spacetime. Conversely one might proceed first by modifying Einstein's theory and address the phenomena that GR does not fully explain, we can consider altering its foundational premises, thereby changing the expected behavior of gravity. Adopting the viewpoint of particle physics, GR is the unique local and Lorentz-invariant theory which can be described as being mediated by a hypothetical particle, the graviton—a massless spin-2 particle which couples to matter \cite{Gupta:1954zz,Weinberg:1965rz,Weinberg:1980kq, Wald:1986bj,Feynman:1996kb}. This theory propagates two degrees of freedom (DOFs).

Lorentz violating \cite{Mattingly:2005re} and non-local gravity \cite{Deser:2007jk} theories have been widely investigated in literature. At the same time, it is possible to maintain Lorentz invariance, given its strong experimental support, as well as locality, 
while instead assuming that the graviton has a finite mass. A theory of massive gravity has been searched since the last century, with Fierz and Pauli \cite{Fierz:1939ix} presenting the first extension to Einstein's theory (for a recent review on massive gravity, see \cite{deRham:2014zqa}). This extension introduced an additional mass term linear in the perturbation around a background metric, resulting in a linearized theory of a massive and self-interacting graviton. However, a non-linear completion of the Fierz-Pauli theory encounters two problems, which are the van Dam–Veltman-Zakharov (vDVZ) discontinuity  \cite{vanDam:1970vg,Zakharov:1970cc} and the Boulware-Deser (BD) ghost problem \cite{Boulware:1972yco, VanNieuwenhuizen:1973fi}. For the latter problem, a theory of a massive spin-2 field, six DOFs exist, one which is a scalar mode, known as the BD ghost. The negative energy associated with this mode leads to an unstable vacuum, rendering the theory pathological. The resolution proposed by de Rham, Gabadaze and Tolley \cite{deRham:2010ik,deRham:2010kj} in their dRGT massive gravity theory involves considering a reference metric that interacts with the GR metric, thereby allowing the presence of an additional fixed spin-2 field. This concept has been explored in 
studies from earlier decades \cite{Rosen:1940zza}. The former problem is related to the fact that a massive spin-2 field propagates five DOFs, so that even in the massless limit one could not restore GR. This was resolved by the Vainshtein mechanism \cite{Vainshtein:1972sx}, where the extra DOF responsible for the vDVZ discontinuity gets screened by its own interactions.

A non-linear extension of the mass term that is invariant under non-linear coordinate transformations was explored in \cite{Arkani-Hamed:2002bjr}, leading to the generalization of the Fierz-Pauli linear mass term found in \cite{deRham:2010ik,deRham:2010kj,Hassan:2011vm}. The dRGT theory massive gravity \cite{deRham:2010ik,deRham:2010kj} is a non-linear ghost-free theory that treats the dynamical metric as in GR interacting with an auxiliary fixed metric. This interaction generates the mass term, and we reduce to GR when in the weak gravity limit \cite{Babichev:2013pfa,Babichev:2010jd,Baccetti:2012bk}. The absence of BD ghost in this theory have been proven in \cite{Hassan:2011hr,Hassan:2011tf,Golovnev:2011aa,Kluson:2012wf,Hassan:2012qv,Kugo:2014hja}. Then, without the BD ghost, dRGT gravity has five DOFs, compared to the two DOFs in the massless case of GR. 


By introducing an auxiliary metric, we arrive to dRGT massive gravity \cite{deRham:2010ik,deRham:2010kj}. We can provide the second metric with dynamic properties by adding a kinetic term for the auxiliary field in the action of massive gravity. The resulting theory is bimetric massive gravity, or bigravity \cite{Hassan:2011zd}, a theory of massive gravity which considers two spin-2 dynamical fields. Similar to massive gravity, the interaction between the metrics gives mass to one field, resulting in one massless field as in GR and one massive spin-2 field. The generalization to multiple interacting spin-2 fields is called multi-metric gravity, or multi-gravity \cite{Hinterbichler:2012cn}. These multi-metric theories are free of BD ghosts up to certain conditions \cite{Wood:2024acv}. They also possess the characteristic of having only one massless graviton, i.e., any additional interacting spin-2 fields should be massive \cite{Boulanger:2000rq}. Thus, bigravity change two assumptions of GR by considering an additional dynamical massive spin-2 fields.

Bimetric gravity is particularly interesting due to its cosmological and astrophysical motivations as it offers a natural resolution to dark matter, arising from the description of interacting spin-2 fields. It serves as viable model for dark matter, incorporating a massive field which interacts gravitationally with the massless mode of GR \cite{Babichev:2016bxi,Babichev:2016hir,GonzalezAlbornoz:2017gbh,Marzola:2017lbt}. Furthermore, bigravity can account for a phase of
late-time acceleration of the universe \cite{deRham:2014fha, Hassan:2014gta,Comelli:2015pua}.



Dualities may also contribute to a better grasp of gravity by enabling us to further explore its properties. For example, AdS/CFT correspondence \cite{Maldacena:1997re}, encodes gravity in a lower dimensional gauge theory. On the other hand, the double copy \cite{Bern:2008qj, Bern:2010yg, Bern:2010ue,Monteiro:2014cda}, is a mathematical connection between color and kinematic factors in the scattering amplitudes of gravitational and gauge theories and is based on the Bern-Carrasco-Johansson (BCJ) duality. The canonical example which manifests this BCJ duality is between (super) Yang-Mills and (super) gravity. Both of these dualities are useful for calculations and for increasing our understanding of gravity. 

In this work, we will focus on classical double copy relations and the bigravity theory. A classical realization of the double copy relations was provided in \cite{Monteiro:2014cda} by considering a family of solutions in GR, namely the Kerr-Schild (KS) solutions \cite{Kerr:1965wfc, Gurses:1975vu, Stephani:2003tm}. This is referred to as the Kerr-Schild double copy. This non-trivial link between two different kind of theories can be studied at the level of the classical solutions, where the solutions in the gravity theory are related with a solution of a gauge theory. Another example of this is the Weyl double copy \cite{Luna:2018dpt,Godazgar:2020zbv}.

In \cite{Monteiro:2014cda}, for example, the Schwarzschild black hole is related to the Coulomb point-like solution via the Kerr-Schild double copy procedure, whereas in \cite{Luna:2016hge}, it was pointed that the formal double copy of the Coulomb solution is any member of the Janis, Newman and Winicour (JNW) solution \cite{Janis:1968zz}. The work in \cite{Monteiro:2014cda} was generalized in \cite{Carrillo-Gonzalez:2017iyj} for maximally symmetric backgrounds, and in \cite{Garcia-Compean:2024uie}, we applied the same formalism in bigravity, implementing the method outlined in \cite{Ayon-Beato:2015qtt}, thereby obtaining the KS double copy relations for certain solutions in bigravity. The implications of massive theories in the double copy relations, even at the scattering level, are not yet fully understood. Improving our understanding of the double copy relations for massive theories at the perturbative and non-perturbative level, is crucial for unlocking new avenues in theoretical physics and refining our comprehension of the interplay between gravity and gauge theories.

This work is organized as follows: In Sec. \ref{Bigravity} we will give a review of the bimetric theory of massive gravity and will provide some comments on how to couple matter to the metrics in bigravity. In Sec. \ref{Double copy} we give a brief summary of the double copy relations and proceed with presenting the classical Kerr-Schild double copy in GR. In Sec. \ref{DKS ansatz} we present the framework for the generalized double Kerr-Schild ansatz in bigravity, followed by time-dependent single Kerr-Schild solutions in Sec. \ref{AdS waves coupled to matter in BG} and novel Pleba\'nski-Demia\'nski solutions in bigravity in Sec. \ref{PD in Bigravity}. Finally, Sec. \ref{FR} is devoted to give our conclusions and final remarks.  

\section{Overview of bigravity} \label{Bigravity}
In this section we briefly overview the theory of bigravity. We will not be exhaustive and we do not attempt to give a complete and detailed analysis. We only will take the opportunity to introduce some notation and conventions which will be necessary for future reference. 
\subsection{Theory of bimetric gravity}
Bigravity theory \cite{Hassan:2011zd} is a massive theory of gravity, incorporating a massive and a massless spin-2 fields which interact gravitationally via a non-derivative potential. The action for bigravity can be expressed in the following form:
{\small\begin{align}
    \label{action_big}
    S_{bi}[g, f]= & \frac{{M^2_{g}}}{2} \int d^{4} x \sqrt{-g} \, R[g]+\frac{{M^2_{f}}}{2} \int d^{4} x \sqrt{-f} \mathcal{R}[f] -m^{2} {M^2_{\text{eff}}}\int d^{4} x \sqrt{-g} \, \mathcal{U}[g, f],
\end{align}}where $M_{g}$ and $M_{f}$ are the Planck masses for each of the metrics $g_{\mu\nu}$ and $f_{\mu\nu}$ which, in natural units, are related to the gravitational constants for each metric by $8\pi G_{g}={M_g}^{-2}$ and $8\pi G_{f}={M_{f}}^{-2}$. The effective Planck mass is given by $M_{\text{eff}} = (M_{g}M_{f})/({M_{g}^{2}+M_{f}^2})^{\frac{1}{2}}$. We can rewrite (\ref{action_big}) in terms of the gravitational couplings $\kappa_{g}$ and $\kappa_{f}$ for each of the metrics, which are related to the gravitational constants as ${\kappa^2_{g}}=16\pi G_{g}$ and ${\kappa^2_{f}}=16\pi G_{f}$. Through this text, we use standard letters, e.g., $R_{\mu\nu}$, for tensors constructed with the $g_{\mu\nu}$ metric and cursive letters such as $\mathcal{R}_{\mu\nu}$ will be used when defining quantities using the $f_{\mu\nu}$ metric. The signature convention we will be using is $(-,+,+,+)$. 

The interaction potential $\mathcal{U}[g,f]$ is written in terms of powers of the interaction matrix $\gamma^{\mu}{}_{\nu}$, given by:
\begin{align}
    \label{int_gamma}
    \mathcal{U}[g,f] & =\sum_{k=0}^{4} b_{k} \mathcal{U}_{k}(\gamma) \, , \qquad \gamma^{\mu}{}_{\nu}=\sqrt{g^{\mu\alpha}f_{\nu\alpha}} \, ,
\end{align}
where $b_{k}$ are coupling constants. The functions $\mathcal{U}_{k}(\gamma)$ in the potential can be written as: 
\begin{align}
    \nonumber \mathcal{U}_{0}(\gamma)&=1\, ,  \quad 
    \mathcal{U}_{1}(\gamma)=[\gamma]\, , \quad 
    \mathcal{U}_{2}(\gamma)=\frac{1}{2 !}\left([\gamma]^{2}-[\gamma^{2}]\right)\,, \\
    \nonumber \mathcal{U}_{3}(\gamma)&=\frac{1}{3 !}\left([\gamma]^{3}-3[\gamma][\gamma^{2}]+2[\gamma^{3}]\right)\, , \\
    \nonumber \mathcal{U}_{4}(\gamma)&=\frac{1}{4 !}\left([\gamma]^{4}-6[\gamma]^{2}[\gamma^{2}]+8[\gamma][\gamma^{3}]+3[\gamma^{2}]^{2}-6[\gamma^{4}]\right) \, , 
\end{align}
where $(\gamma^{n})^{\mu}{}_{\nu}=\gamma^{\mu}{}_{\alpha_{1}}\gamma^{\alpha_{1}}{}_{\alpha_{2}}\cdots \gamma^{\alpha_{n-1}}{}_{\nu}$ and the trace of the matrix $\gamma^{\mu}{}_{\nu}$ is defined as $[\gamma]\equiv \operatorname{tr}(\gamma^{\mu}{}_{\nu})=\gamma^{\mu}{}_{\mu}\,$. The action in (\ref{action_big}) reduces to the Fierz-Pauli action \cite{Fierz:1939ix} for linear massive gravity when $b_{2} =-1-2b_{3}-b_{4}$ \cite{Hassan:2011vm}. By varying the action (\ref{action_big}) with respect to the fields, we can obtain the equations of motions. Using the gravitational couplings $\kappa_{g}$, $\kappa_{f}$ and $\kappa$, a function of both couplings, the resulting equations of motion are:
\begin{align}
\label{eom_big}
    G^{\mu}{ }_{\nu}=\frac{m^{2} {\kappa_{g}}^{2}}{{\kappa}^{2}} V^{\mu}{ }_{\nu}\equiv  Q^{\mu}{}_{\nu} \,, \qquad \ \ \ \ \ \ \mathcal{G}^{\mu}{}_{\nu}=\frac{m^{2} {{\kappa_{f}}^{2}}}{{\kappa}^{2}} \mathcal{V}^{\mu}{ }_{\nu} \equiv  \mathcal{Q}^{\mu}{}_{\nu} \, .
\end{align}
$G^{\mu}{}_{\nu}$ and $\mathcal{G}^{\mu}{}_{\nu}$ are the Einstein tensors for the metrics $g_{\mu\nu}$ and $f_{\mu\nu}$ respectively. The quantities $V^{\mu}{}_{\nu}$ and $\mathcal{V}^{\mu}{}_{\nu}$ are defined as:
\begin{align}
    \nonumber V^{\mu}{}_{\nu} \equiv \frac{2 g^{\mu \alpha}}{\sqrt{-g}} \frac{\delta(\sqrt{-g} \, \mathcal{U})}{\delta g^{\alpha \nu}}=\tau^{\mu}{ }_{v}-\mathcal{U} \delta^{\mu}{ }_{\nu} \, , \quad \mathcal{V}^{\mu}{}_{\nu}  \equiv \frac{2 f^{\mu \alpha}}{\sqrt{-f}} \frac{\delta(\sqrt{-g} \, \mathcal{U})}{\delta f^{\alpha \nu}}=-\frac{\sqrt{-g}}{\sqrt{-f}} \tau^{\mu}{ }_{\nu},
\end{align}
and they are effective energy-momentum tensors arising from the interaction between the two metrics. The defined quantities $Q^{\mu}{}_{\nu}$ and $\mathcal{Q}^{\mu}{}_{\nu}$ will be referring to as the interaction tensors. Moreover in the previous equations $\tau^{\mu}_{ \ \nu}$ is given by
\begin{align}
    \nonumber \tau^{\mu}{}_{\nu} \equiv & \left(b_{1} \mathcal{U}_{0}+b_{2} \mathcal{U}_{1}+b_{3} \mathcal{U}_{2}+b_{4} \mathcal{U}_{3}\right) \gamma^{\mu}{ }_{\nu} -\left(b_{2} \mathcal{U}_{0}+b_{3} \mathcal{U}_{1}+b_{4} \mathcal{U}_{2}\right)(\gamma^{2})^{\mu}{ }_{\nu} \\
    \nonumber & +\left(b_{3} \mathcal{U}_{0}+b_{4} \mathcal{U}_{1}\right)(\gamma^{3})^{\mu}{ }_{\nu} -b_{4} \mathcal{U}_{0}(\gamma^{4})^{\mu}{ }_{\nu} \, . 
\end{align}
We will discuss a couple of options to coupling matter to the fields in bigravity shortly, but for the sake of writing the equations of motion for the metrics, we consider some matter Lagrangians $\mathcal{L}^{g}_{M}$ and ${\mathcal L}^{f}_{M}$ which describe the matter content coupled to their respective metric. Thus they will describe decoupled matter sectors in the sense they are coupled only to $g_{\mu\nu}$ and to $f_{\mu\nu} $ respectively. By definition, we can vary these Lagrangians to obtain the energy-momentum tensors ${T_{M}}^{\mu}{}_{\nu}$ and ${\mathcal{T}_{M}}^{\mu}{}_{\nu}$:
\begin{align}
    \nonumber {T_{M}}_{\mu \nu} \equiv\frac{-2}{\sqrt{-g}}\frac{\delta}{\delta g^{\mu \nu}}\left(\sqrt{-g}\,\mathcal{L}_{M}^{g}\right) \, , \qquad
    \nonumber 
    {\mathcal{T}_{M}}_{\mu \nu} \equiv \frac{-2}{\sqrt{-f}}\frac{\delta}{\delta f^{\mu \nu}}\left(\sqrt{-f}\,{\mathcal L}^{f}_{M}\right) \, .
\end{align}
Then, we can rewrite the equations of motion of bigravity by defining a couple of traces and trace-reversed tensors, such as the interaction tensors $\check{Q}^{\mu}{}_{\nu}$ and $ \mathcal{\check{Q}}^{\mu}{}_{\nu}$ and the energy-momentum tensors ${\check{T}_{M}{}}^{\mu}{}_{\nu}$ and $ {\check{\mathcal{T}_M}}^{\mu}{}_{\nu}$ as follows:
\begin{align}
     \nonumber  \check{Q}^{\mu}{}_{\nu} &\equiv \, Q^{\mu}{}_{\nu} - \frac{1}{2}Q\, \delta^{\mu}{}_{\nu}   \,, \quad \check{\mathcal{Q}}^{\mu}{}_{\nu} \equiv \mathcal{Q}^{\mu}{}_{\nu} - \frac{1}{2}\mathcal{Q} \, \delta^{\mu}{}_{\nu} \, , \quad \\
     \nonumber  {\check{T}_{M}{}}^{\mu}{}_{\nu} &\equiv \,  {T_{M}}^{\mu}{}_{\nu}- \frac{1}{2}T_{M}  \, \delta^{\mu}{}_{\nu} \, , \quad {\check{\mathcal{T}_M}}^{\mu}{}_{\nu} \equiv {\mathcal{T}_{M}}^{\mu}{}_{\nu}- \frac{1}{2}\mathcal{T}_{M} \, \delta^{\mu}{}_{\nu} \,,  \quad 
\end{align}
where $Q\equiv Q^{\mu}{}_{\mu}$, $\mathcal{Q}\equiv \mathcal{Q}^{\mu}{}_{\mu}$, $T_{M} \equiv {T_{M}}^{\mu}{}_{\mu}$ and $\mathcal{T}_{M} \equiv {\mathcal{T}_{M}}^{\mu}{}_{\mu}$,

Using these definitions, the trace-reversed equations of bigravity coupled to matter are respectively:
\begin{align}
\label{trace_rev_matter}
    R^{\mu}{}_{\nu} - \check{Q}^{\mu}{}_{\nu} = \frac{{\kappa_{g}}^{2}}{2} {\check{T}_{M}{}}^{\mu}{}_{\nu} \,, \qquad \mathcal{R}^{\mu}{}_{\nu} - \check{\mathcal{Q}}^{\mu}{}_{\nu}= \frac{{\kappa_{f}}^{2}}{2} {\check{\mathcal{T}_M}}^{\mu}{}_{\nu} \, . 
\end{align}
We can couple different types of matter to the metrics in different manners, which will be discussed briefly. The equations in (\ref{trace_rev_matter}) apply when the metrics $g_{\mu\nu}$ and $f_{\mu\nu}$ are coupled independently to matter or in some other way. In the latter case, the form of ${T_{M}}^{\mu}{}_{\nu}$ and ${\mathcal{T}_{M}}^{\mu}{}_{\nu}$ will be different from the example presented here, but we will have a energy-momentum tensor originating from matter content coupled to the corresponding metrics.

\subsection{Coupling matter to bigravity}

One important point is to determine how the matter will be coupled to bigravity. For example, is possible to couple some matter content ${T_{M}}_{\mu\nu}$ to the metric $g_{\mu\nu}$ and, independently to this coupling, and to couple ${\mathcal{T}_{M}}_{\mu\nu}$ to the metric $f_{\mu\nu}$, which would be described by
\begin{align}
    \nonumber S_{\text{bi,M}}[g,f,\text{Matter}]= \,& S_{\text{bi}}[g,f]+S_{\text{Mat}}[g,f,\text{Matter}] \, ,\\ 
    \nonumber S_{\text{Mat}}[g,f,\text{Matter}]\equiv \, & S_{\text{Mat}}[g,\ldots] +  S_{\text{Mat}}[f, \ldots] \\
    = & \, \int\text{d}^{4}x\sqrt{-g}\, \mathcal{L}^{g}_{M}\left({g}_{\,\mu \nu},\ldots\right) + \int\text{d}^{4}x\sqrt{-f}\, \mathcal{L}^{f}_{M}\left({f}_{\,\mu \nu},\ldots\right) \, ,
\end{align}
where '$\ldots$' represent the matter fields coupled to each metric. This important problem was worked out in detail in Ref. \cite{Ayon-Beato:2015qtt}. This case was mentioned before, and the matter content of each metric does not interact with each other, remaining confined to their respective metric. One example of this case is presented in \cite{Ayon-Beato:2015qtt}, where a Kerr-Newmann-(A)dS in one metric is interacting with a Kerr-(A)dS solution in the other. Here, one metric is coupled to the Maxwell electromagnetic tensor ${T_{M}}_{\mu\nu}$ and for the other metric the energy-momentum tensor is zero, ${\mathcal{T}_{M}}_{\mu\nu}=0$. We can also have the case where ${\mathcal{T}_{M}}_{\mu\nu} \ne 0$ and only couples to $f_{\mu\nu}$, which has a similar behaviour or we can also have electrovacuum in both metrics. This can be seen as one possibility on how to couple matter to the metrics in bigravity.

In the following sections we will consider examples where we have matter coupled to both metrics but not in an independent way as was mentioned before. We will consider the case where the matter content is symmetrically coupled to the metric fields and study the results in the context of the double copy. This symmetric coupling is provided by the effective metric presented in the next part.

\subsubsection{Coupling matter via the effective metric}
We can consider the case where the two metrics $g_{\mu\nu}$ and $f_{\mu\nu}$ are coupled to the same external matter source and distribute the matter coupling in a symmetric form, giving equal relevance the two metrics in bigravity. We can carry out this coupling by using the effective metric, where the action of the theory when coupled to matter is:
\begin{align}
    \label{mat_bigrav}
    S_{\text{bi,M}}[g,f,\text{Matter}]=S_{\text{bi}}[g,f]+\int\text{d}^{4}x\sqrt{-g_{E}}\, \mathcal{L}_M\left({g_{E}}_{\,\mu \nu},\ldots\right),
\end{align}
with $\mathcal{L}_M$ the Lagrangian for the matter, which depends on the matter content and the effective metric \cite{deRham:2014naa}, which is defined as
\begin{align}
    \label{eff_metric}
    {g_{E}}_{\,\mu\nu}\equiv \alpha^{2}g_{\mu\nu}+2\alpha\beta g_{\mu\lambda}\,\gamma^{\lambda}{}_{\nu}+\beta^2f_{\mu\nu}\, .
\end{align}
The coupling is ``symmetric" in the sense that the metric ${g_{E}}_{\,\mu\nu}$ is symmetric under simultaneous exchange of the metric fields  $g_{\mu\nu} \leftrightarrow f_{\mu\nu}$ and the coupling constants $\alpha \leftrightarrow \beta$ in the usual case where the vierbein and metric formalism coincide. The energy-momentum tensor for each metrics can be written  as:
\begin{align}
    \nonumber {T_{M}}_{\mu \nu} \equiv\frac{-2}{\sqrt{-g}}\frac{\delta}{\delta g^{\mu \nu}}\left(\sqrt{-g_{E}}\,\mathcal{L}_{M}\right)=\frac{\sqrt{-g_{E}}}{\sqrt{-g}}\,{T_{E}}_{\rho\sigma}\frac{\delta {g_{E}}^{\rho\sigma}}{\delta g^{\mu \nu}}\, , \\
    \nonumber 
    {\mathcal{T}_{M}}_{\mu \nu} \equiv \frac{-2}{\sqrt{-f}}\frac{\delta}{\delta f^{\mu \nu}}\left(\sqrt{-g_{E}}\,{\mathcal L}_{M}\right)=\frac{\sqrt{-g_{E}}}{\sqrt{-f}}\,{T_{E}}_{\rho\sigma}\frac{\delta {g_{E}}^{\rho\sigma}}{\delta f^{\mu \nu}}\, ,
\end{align}
where we have rewritten the matter sources ${T_{M}{}}^{\mu}{}_{\nu}$ and ${\mathcal{T}_{M}{}}^{\mu}{}_{\nu}$ for the metrics $g_{\mu\nu}$ and $f_{\mu\nu}$ respectively in terms of the energy-momentum tensor ${T_{E}}_{\mu\nu}$, which will be defined by the effective metric ${g_{E}}_{\mu\nu}$.

We are interested for generalized Kerr-Schild metrics, and for such family of solutions, the effective metric and its inverse reduce to:
\begin{align}
    \nonumber {g_{E}}_{\,\mu\nu}=(\alpha+\beta C)\left(\alpha g_{\mu\nu}+\frac{\beta}{C}f_{\mu\nu}\right)\, , \quad {g_{E}{}}^{\mu\nu}=\frac{1}{\left(\alpha+\beta C\right)^{3}}\left(\alpha g^{\mu\nu}+\beta C^{3}f^{\mu\nu}\right),  
\end{align}
which implies that the energy-momentum tensors are related as follows
\begin{align}
\label{eff em tensors}
    \frac{1}{\alpha}{T_{M}}_{\mu\nu}=\frac{C}{\beta}{\mathcal{T}_{M}}_{\mu\nu}=(\alpha+\beta\,C){T_{E}}_{\mu\nu}\,.
\end{align}
With this machinery we can consider some examples in bigravity where both metrics are coupled to external matter, in order to study them in the context of the double copy. Before proceeding with the examples, we will consider the perturbative double copy and the Kerr-Schild double copy, and some comments about massive gravity in this context.

\section{Double copy relations in GR and massive gravity}\label{Double copy}

In this section we will discuss the classical double copy in the Kerr-Schild approach and comment about the double copy of massive theories. 

\subsection{Brief overview of the Kerr-Schild double copy}
The double copy starting as a mathematical correspondence between scattering amplitudes in gravity and gauge theories \cite{Bern:2008qj, Bern:2010yg, Bern:2010ue}. This correspondence links the amplitudes of both theories by using a correspondence between color and kinematic factors that appear in the amplitudes of both kinds of processes. This relation between numerators of the scattering amplitudes is known as the Bern-Carrasco-Johansson (BCJ) duality. One known example with manifest BCJ duality was studied in \cite{Bern:2010ue} and relates (super) Yang-Mills and (super) gravity. In this example, with the BCJ duality being manifest, one can construct the tree-level amplitudes of gravity processes by a product of the gauge theory amplitudes. As expressed in \cite{Bern:2022wqg}, we can start from the tree-level gauge theory amplitudes $\mathcal{A}_m^{\text{tree}}$ and construct the tree-level gravity amplitudes $\mathcal{M}_m^{\mathrm{tree}}$ by the substitutions
\begin{align*}
    c_{i} \rightarrow \widetilde{n}_i \, , \quad  g \rightarrow \frac{\kappa}{2} \, , \qquad \qquad n_{i} \rightarrow \widetilde{c}_i \, , \quad  g \rightarrow y \, ,
\end{align*}
where the first set of substitutions correspond to the double copy procedure and replaces a color factor $c_{i}$ of the gauge theory for with a Lorentz factor $\widetilde{n}_{i}$ and the coupling constant of the gauge theory $g$, with the coupling constant of the gravitational theory, $\kappa$. The second set of substitutions correspond to the zeroth copy procedure are given for completeness, and $y$ as the coupling constant of a scalar theory. Then, by doing the double copy procedure we can construct $\mathcal{M}_m^{\mathrm{tree}}$ using $\mathcal{A}_m^{\text{tree}}$ as: 
\begin{align*}
    i\mathcal{A}_m^{\text{tree}}=g^{m-2}\sum_i\frac{c_in_i}{\prod_{i_j}D_{i_j}}\, \quad \Rightarrow \quad 
    i\mathcal{M}_m^{\mathrm{tree}} =\mathcal{A}_m^{\text{tree}}\bigg|_{{\substack{c_i\to\check{n}_i \\ g\to\frac{\kappa}
    {2}}}} =\left(\frac{\kappa}{2}\right)^{m-2}\sum_i\frac{\tilde{n}_i n_i}{\prod_{i_j}D_{i_j}} \, . 
\end{align*}
We have a similar expression for the zeroth copy and for the L-loop level amplitudes, where the double copy relations are conjectured to hold  \cite{Bern:2022wqg}. 
Here, $D_{i_j} \propto p^{2}_{i_{j}}$, for massless particles, which is the case for the gluons. The index $i\in \Gamma$, runs over cubic graphs, and the index $j$ considers the propagators of each graph.

In order to deepen the understanding on the double copy, there has also been interest in the case where we have massive fields \cite{Johansson:2019dnu, Plefka:2019wyg}, as the implication of considering massive fields in the double copy procedure are not been entirely clarified. There has been work regarding the double copy and massive theories, which can be found in \cite{Bern:2022wqg} and which connects massive gravity with spontaneously broken Yang-Mills theory \cite{Johnson:2020pny,Momeni:2020vvr}. 

For massive gravity and the double copy, the authors \cite{Momeni:2020vvr} tried to relate de Rham-Gabadadze-Tolley (dRGT) massive gravity, where $f_{\mu\nu}$ is not dynamical, as a double copy of massive Yang-Mills theory. The decoupling limits of Yang-Mills theory and dRGT massive gravity are the NLSM and special Galileon respectively, and is noted that the decoupling limit and double copy procedures do not commute. Also, while the 3 and 4 point massive Yang-Mills amplitudes give massive gravity amplitudes, the relation fails at higher points, due to the appearance of spurious poles. Studying the double copy of massive gravity theories can help us gain understanding regarding this duality. 

As commented in \cite{Bern:2019prr}, introducing sources in the gravity side can be confusing. This is due to the fact that adding a source in the gravity theory adds a source in the gauge theory which may break gauge invariance and that we also need to make a specific choice source which may itself receive corrections order by order in the perturbative theory, as discussed in \cite{Duff:1973zz,sardelis1975tree}. Even if this is the case for the scattering amplitudes story when coupling matter to the gravity theory, we can study which classical equations the fields follow, as we will do by coupling bigravity to external matter. 

With all these considerations in mind, we are motivated to study the classical double copy of a massive gravity theory, such as bimetric gravity. In particular, as has been studied in GR in \cite{Luna:2015paa,Farnsworth:2023mff}, we will consider the double Kerr-Schild family of solutions for bigravity in maximally symmetric backgrounds and obtain the equations of motion for the double, single and zeroth copy fields.

\subsection{Kerr-Schild double copy in GR}

The double copy relations and BCJ duality were discovered at the level of scattering amplitudes of gauge and gravity theories. This is a perturbative approach, and we may be interested in considering if there is a non-perturbative realization of these mathematical relations. Some approaches to this idea are the Kerr-Schild double copy \cite{Monteiro:2014cda}, the Weyl double copy \cite{Luna:2018dpt,Godazgar:2020zbv}, where  the linearized Weyl curvature tensor is related to the linearized Yang-Mills field strength tensor, and the three-dimensional analog of this relation is the Cotton double copy \cite{CarrilloGonzalez:2022mxx}, and here we will be consider the former one. Recently it was found that the double copy in $AdS_3$ can be related to a minitwistor space and some implications for the Cotton double copy of waves and black holes in topologically massive gravity were studied \cite{Beetar:2024ptv}. 

The double copy relates gravitational theories to the square of gauge theories. While it is well understood in flat backgrounds, its precise realization around curved spacetimes remains an open question. In this paper, we construct a classical double copy for cohomology class representatives in the minitwistor space of hyperbolic spacetimes. We find that the realisation of a physical double copy requires that the masses of the different spinning fields are not equal, contrary to the flat space prescription. This leads to a position-space double copy for bulk-to-boundary propagators. We also show that in coordinate space, this implies the Cotton double copy for waves and warped black holes of Topologically Massive Gravity. We show that these are exact double copy relations by constructing their Kerr-Schild metrics and also analysing the Kerr-Schild double copy. Furthermore, we find that near the boundary the double copy relates the dual CFT currents.

The Kerr-Schild classical double copy \cite{Monteiro:2014cda} is a procedure which considers that certain classical gauge and gravity solutions can be related by a double copy relation and involves the Kerr-Schild metrics \cite{Kerr:1965wfc, Gurses:1975vu, Stephani:2003tm}, a special family of solutions of the Einstein's equations. In the case of GR, this classical approach was studied in \cite{Monteiro:2014cda} for Minkowski background and in maximally symmetric backgrounds in \cite{Carrillo-Gonzalez:2017iyj}. The Kerr-Schild ansatz in curved backgrounds has the following form:
\begin{align}
    \label{ks_1}
    g_{\mu\nu}= \overline{g}_{\mu\nu}+\kappa h_{\mu\nu} \, ,  \quad h_{\mu\nu} = \phi k_{\mu}k_{\nu} \, , 
\end{align}
where $\kappa^{2}=16\pi G_{N}$ is the gravitational coupling, $G_{N}$ is the Newton's constant and $\overline{g}_{\mu\nu}$ is the background metric. We have also defined the perturbation to the background metric, $h_{\mu\nu}$, where  $\phi$ is a scalar function and $k_{\mu}$ is a vector which is null with respect to the two metrics and geodesic with respect to $g_{\mu\nu}$, so it is also geodesic with respect to $\overline{g}_{\mu\nu}$ \cite{Stephani:2003tm}. From the ansatz (\ref{ks_1}), the Ricci tensor can be written as
\begin{align}
    \label{ric_1}
    R^{\mu}{}_{\nu}=\overline{R}^{\mu}{}_{\nu}-\kappa\, h^{\mu\lambda} \overline{R}_{\lambda \nu}+\frac{\kappa}{2}\overline{\nabla}_{\lambda}\left[\overline{\nabla}^{\mu}h^{\lambda}{}_{\nu}+\overline{\nabla}_{\nu}h^{\mu\lambda}-\overline{\nabla}^{\lambda}h^{\mu}{}_{\nu}\right] \ 
\end{align}
where $R^{\mu}{}_{\nu}=g^{\alpha\mu}R_{\alpha\nu}$. The specific combination of indices in (\ref{ric_1}) linearizes the Ricci tensor in the field $h_{\mu\nu}$. $\overline{\nabla}_{\mu}$ is the covariant derivative compatible with the background metric $\overline{g}_{\mu\nu}$ via $\overline{\nabla}_{\lambda}\, \overline{g}_{\mu\nu}=0$, with $\overline{\nabla}^{\mu}=\overline{g}^{\mu\alpha}\overline{\nabla}_{\alpha}$, and the same for $\nabla_{\mu}$ which is associated to $g_{\mu\nu}$, i.e., $\nabla_{\lambda}\, g_{\mu\nu}=0 \,$. Using Einstein's equations and (\ref{ric_1}), one can obtain the equations of motion for the single and zeroth copy fields as in \cite{Carrillo-Gonzalez:2017iyj} for maximally symmetric backgrounds, i.e., spacetimes with constant curvature, such as flat, de Sitter or anti-de Sitter backgrounds, which correspond to zero, positive and negative scalar curvature, so that the fields defined in the Kerr-Schild double copy, $A_{\mu} = \phi k_{\mu}$ and the scalar function $\phi$ satisfy the equations:
\begin{align*}
     \overline{\nabla}_{\lambda}F^{\lambda\mu} = \, 0 \,, \quad \overline{\nabla}^{2}\phi-\frac{\overline{R}}{6}\phi = \, 0 \,  \qquad \xrightarrow{\overline{g}_{\mu\nu}=\eta_{\mu\nu}} \qquad  \nonumber \partial_{\mu}F^{\mu\nu}=0 \, , \quad \partial^{2}\phi =0\, ,
\end{align*}
with $F_{\mu\nu}\equiv \overline{\nabla}_{\mu}A_{\nu}- \overline{\nabla}_{\nu}A_{\mu}$. Thus fields $A_{\mu}$ obeys a Maxwell equation and $\phi$ a Klein-Gordon equation, both in a curved background and which reduces to \cite{Monteiro:2014cda} in the flat background case, i.e., when $\overline{g}_{\mu\nu}=\eta_{\mu\nu}$. 

Some examples of stationary solutions include the Schwarzschild metric and its generalization with a rotation parameter, electric charge, and curved background, which describe black holes with different properties and which allow a Kerr-Schild form. There have been other Kerr-Schild solutions in GR studied in the double copy framework, e.g., AdS waves in \cite{Carrillo-Gonzalez:2017iyj}.

In the classical Kerr-Schild double copy $h_{\mu\nu}$, is related to single copy field, by replacing a Lorentz factor $k_{\mu}$ with a color factor $c^{a}$ to obtain the non-Abelian single copy field $A^{a}_{\mu}$. Similarly, for the zeroth copy, we eliminate another $k_{\mu}$ and add a color factor $c^{a'}$ and obtain the biadjoint scalar field $\Phi^{aa'}$. The fields $A^{a}_{\mu} $ and $ \Phi^{aa^{\prime}}$ with color factors would be the single and zeroth copy fields, but we will be referring to as the color-stripped or Abelian fields $A_{\mu}$ and $\phi$ as the single and zeroth copy fields. The fields $A_{\mu}$ and $\phi$ are equivalently obtained by removing one or two the Lorentz factors $k_{\mu}$ respectively from the graviton $h_{\mu\nu}$. The coupling constants are also replaced, with the gravity theory coupling substituted to the gauge coupling constant for the single copy, and the gauge coupling replacement with the scalar theory coupling in the zeroth copy. This can be summarized with the statement that, in the classical Kerr-Schild double copy, the replacements:
\begin{align}
    \nonumber A^{a}_{\mu}=c^{a}\, \phi\,k_{\mu}\, , \quad \frac{\kappa}{2} \rightarrow g \, , \qquad \Phi^{aa'}=c^{a}\tilde{c}^{a'} \phi\, , \quad g \rightarrow y  \,,
\end{align}
relates our double, single and zeroth copy solutions. Here $y$ is the coupling constant pf the zeroth copy. For the KS ansatz, $A^{a}_{\mu}$ is a solution of the Yang-Mills equations for arbitrary color factors $c^{a}$. Then, due to the linear structure of the Einstein's equations, these fields satisfy the Abelian equations \cite{Monteiro:2014cda}, so that we can work with the color-stripped fields $A_{\mu},\phi$, 

As commented before, the double copy map from gauge to gravity considers the dilaton and the Kalb-Ramond 2-form \cite{Monteiro:2014cda}. The ansatz in (\ref{ks_1}) consider only the symmetric tensor $h_{\mu\nu}$, so that there is no antisymmetric 2-form, and because $h_{\mu\nu}$ is traceless due to the null property of the KS solutions, the dilaton is also absent. Then, this is a double copy map between the gauge theory and pure GR.

As it is remarked in \cite{Luna:2016hge}, the double copy of the Coulomb gauge solution, as in \cite{Kim:2019jwm}, is ambiguous because it is any member of the JNW family of solution \cite{Janis:1968zz}, which includes the Schwarzschild metric and this is because we can choose whether the dilaton is sourced or not after taking the double copy. Then, Schwarzschild solution can be interpreted as the double copy of the Coulomb solution as in certain cases. We will be considering other solutions and its double copy relations.

\subsection{Kerr-Schild double copy in bigravity}

A question that arises is whether it is possible to carry out the classical double copy procedure in theories of massive gravity, as the implications of using scattering amplitudes of massive theories in the double copy are not yet fully understood. Efforts have been made in this direction \cite{Momeni:2020vvr,Johnson:2020pny,Bern:2019prr}, and in order to study the double copy relations in a classical context for massive gravity, in \cite{Garcia-Compean:2024uie} we employed the generalized Kerr-Schild ansatz \cite{Kerr:1965wfc, Gurses:1975vu, Stephani:2003tm}:
\begin{align}
\label{ansatz_1}
    \nonumber g_{\mu\nu}= &\overline{g}_{\mu\nu}+\kappa_{g}h_{\mu\nu} \, ,  \quad h_{\mu\nu} = \phi_{g}k_{\mu}k_{\nu} \, , \\
    f_{\mu\nu}\, = & \, C^{2}(\overline{g}_{\mu\nu}+\kappa_{f}\mathscr{h}_{\mu\nu}) \, , \quad \mathscr{h}_{\mu\nu} = \phi_{f}k_{\mu}k_{\nu} \, , 
\end{align}
motivated by previous work in Ref. \cite{Ayon-Beato:2015qtt}, to reinterpret the Kerr-Schild double copy equations of such solutions in bigravity. In (\ref{ansatz_1}), the full metrics $g_{\mu\nu}, f_{\mu\nu}$ are written in terms of the background metrics $\overline{g}_{\mu\nu}, \overline{f}_{\mu\nu}$ and its corresponding perturbations $h_{\mu\nu},\mathscr{h}_{\mu\nu}$. $C$ is restricted to be a proportionality constant by the Bianchi constraints \cite{Hassan:2012wr} and permit us have some freedom when choosing the cosmological constant for each metric. For proportional metrics $\overline{f}_{\mu\nu}=\overline{g}_{\mu\nu}$ and same null vectors $k_{\mu}$ in both metrics, this leads to a closed form for the interaction matrix (\ref{int_gamma}), as was shown in  \cite{Ayon-Beato:2015qtt}. The vector $k_{\mu}$ is also geodetic with respect to the full metrics. If the contraction of the interaction tensors with the null vector is zero, then it is also geodetic with respect to the background metric \cite{Garcia-Compean:2024uie}. 

In \cite{Garcia-Compean:2024uie} we considered maximally symmetric backgrounds in bigravity and proceeded as in \cite{Carrillo-Gonzalez:2017iyj} to obtain the equations of motion time-dependent wave solutions and stationary black holes, at the level of the classical double, single and zeroth copy. For the black hole stationary solutions in bigravity, we obtained
\begin{align}
    \nonumber  \overline{\nabla}_{\lambda}F^{\lambda\mu} = \, 0 \, , \quad \overline{\nabla}^{2}\phi_{g}-\frac{\overline{R}}{6}\phi_{g} = \, 0 \, ; \qquad
    \overline{\nabla}_{\lambda}\mathcal{F}^{\lambda\mu} = \,  0 \, ,  \quad  \,^{(f)}\overline{\nabla}^{2}\phi_{f}-\frac{\overline{\mathcal{R}}}{6}\phi_{f}  =  \,0 \, ,
\end{align}
where the strength tensors are defined in terms of their respective vector field. This is similar to what is found in GR, and we note that the equations of motion are decoupled. The case is different for time-dependent solutions, such as bigravitational waves \cite{Ayon-Beato:2018hxz}, where the two perturbations contribute to both equations of motion. 

The results obtained by contracting the gravity equations of motion with Killing vectors can be interpreted in the Kerr-Schild double copy context \cite{Monteiro:2014cda, Carrillo-Gonzalez:2017iyj} as done in for bigravity in \cite{Garcia-Compean:2024uie}. By replacing one of the null vectors $k_{\mu}$ from the spin-2 fields $h_{\mu\nu}$ and $\mathscr{h}_{\mu\nu}$, with color factors $c^{a}$ and $\mathscr{c}^{a}$ from the gauge theory, one can obtain the non-Abelian fields $A^{a}_{\mu},\mathcal{A}^{a}_{\mu}$ which satisfy their respective Yang-Mills equations. By the same procedure, the zeroth copy fields, ${\Phi_{g}}^{aa'},{\Phi_{f}}^{aa'}$ may be obtained, so that these can be expressed as:
\begin{align}
\label{replace}
    \nonumber A^{a}_{\mu}=c^{a}\, \phi_{g}\,k_{\mu}\, , \quad \frac{\kappa_{g}}{2} \rightarrow g_{g} \, , \qquad \mathcal{A}^{a}_{\mu}=\mathscr{c}^{a}\, \phi_{f}\,k_{\mu} \,, \quad \frac{\kappa_{f}}{2} \rightarrow g_{f} \, ,\\
    {\Phi_{g}}^{aa'}=c^{a}\tilde{c}^{a'} \phi_{g}\, , \quad g_{g} \rightarrow y_{g} \, , \qquad {\Phi_{f}}^{aa'}=\mathscr{c}^{a}\tilde{\mathscr{c}}^{a'}\phi_{f}\,, \quad g_{f} \rightarrow y_{f} \, ,
\end{align}
Apart from changing kinematic to color factors or viceversa,  the coupling constants are also substituted. $\kappa_{g}, \kappa_{f}$ are replaced with the gauge theory coupling constants $g_{g}, g_{f}$ in the single copy, as well as any gravity source to a color source, and a similar procedure for the zeroth copy, where $y_{g}$ and $y_{f}$ are the coupling constants for the biadjoint theory. In (\ref{replace}), we use the same null vectors for both metrics, $l_{\mu}=k_{\mu}$, due to the choice of ansatz (\ref{ansatz_double ks bg}) and as in \cite{Monteiro:2014cda}, one can work with the Abelian version of this fields, i.e., $A_{\mu},\mathcal{A}_{\mu}$ and $\phi_{g},\phi_{f}$ without the color factors, which we will refer to as the single and zeroth copy fields.

Once the formalism of the Kerr-Schild ansatz is extended to massive gravity solutions, we can think one extending the family of solutions to a double Kerr-Schild formalism. In \cite{Luna:2015paa}, the KS double copy of Taub-NUT solution was presented, obtaining that the single copy of such solution is a dyon gauge theory. This Taub-NUT soltion was given in a double Kerr-Schild form, which allowed a KS double copy procedure. In this same line, we are interested in extending the generalized KS solutions for bigravity studied in \cite{Garcia-Compean:2024uie} in the context of the double copy by working with double Kerr-Schild solutions of this massive gravity theory.

\section{Double Kerr-Schild ansatz in bigravity and the double copy}\label{DKS ansatz}

In this section we extend the work done in \cite{Garcia-Compean:2024uie} by considering a larger family of solutions, which can be written in a double Kerr-Schild ansatz. We will work with Pleba\'nski-Demia\'nski stationary double Kerr-Schild solutions, which generalize Kerr-Newman solutions by considering some additional parameters. Thus, our main goal is to obtain the double copy equations of motion for this solution in bigravity. 

Let us consider a double Kerr-Schild ansatz for both metrics:
\begin{align}
\label{ansatz_double ks bg}
    \nonumber g_{\mu\nu} &= \overline{g}_{\mu\nu}+\kappa_{g}\,h_{\phi\,\mu\nu}+\kappa_{g}\, h_{\psi\,\mu\nu} \, ,  \quad &h_{\phi\,\mu\nu} = \phi_{g}k_{\mu}k_{\nu} \, , \quad  h_{\psi\, \mu\nu} &= \psi_{g}l_{\mu}l_{\nu} \\
    f_{\mu\nu}\, &=  \, C^{2}(\overline{f}^{*}_{\mu\nu}+\kappa_{f}\mathscr{h}_{\phi\, \mu\nu} + \kappa_{f}\mathscr{h}_{\psi\, \mu\nu}) \, , \quad &\mathscr{h}_{\phi\, \mu\nu} = \phi_{f} \mathscr{k}_{\mu}\mathscr{k}_{\nu} \, , \quad \mathscr{h}_{\psi\, \mu\nu} &= \psi_{f} \mathscr{l}_{\mu}\mathscr{l}_{\nu}  \, , 
\end{align}
where $\overline{f}^{}_{\mu\nu}=C^{2}\overline{f}^{*}_{\mu\nu}$. We also write the full perturbations as the sum:
\begin{align}
    \label{full_pert}
    \nonumber & h_{\mu\nu}=h_{\phi\,\mu\nu}+ h_{\psi\,\mu\nu} \, , \quad \mathscr{h}_{\mu\nu}=\mathscr{h}_{\phi\,\mu\nu}+ \mathscr{h}_{\psi\,\mu\nu}\, , \\
    \Rightarrow \quad & g_{\mu\nu} = \overline{g}_{\mu\nu}+\kappa_{g}\,h_{\mu\nu} \, , \quad 
    f_{\mu\nu}\, =  \, C^{2}(\overline{f}^{*}_{\mu\nu}+\kappa_{f}\mathscr{h}_{ \mu\nu})\, . 
\end{align}
The vectors $k_{\mu}$, $l_{\mu}$ and $\mathscr{k}_{\mu}$, $\mathscr{l}_{\mu}$ are null and geodetic with respect to $g_{\mu\nu}$ and $f_{\mu\nu}$ respectively, and also null with respect to their background metrics:
\begin{align*}
    g^{\mu\nu} k_{\mu} k_{\nu} = \overline{g}^{\mu\nu} k_{\mu} k_{\nu} &= 0 = g^{\mu\nu} l_{\mu} l_{\nu} = \overline{g}^{\mu\nu} l_{\mu} l_{\nu} \,, \quad &k^{\mu} \,^{(g)}\nabla_{\mu} k_{\nu}  = 0 \, , \quad l^{\mu} \,^{(g)}\nabla_{\mu} l_{\nu} &= 0,\\
    f^{\mu\nu}\mathscr{k}_{\mu} \mathscr{k}_{\nu} = \overline{f}^{\mu\nu} \mathscr{k}_{\mu} \mathscr{k}_{\nu} &= 0 = f^{\mu\nu} \mathscr{l}_{\mu} \mathscr{l}_{\nu} = \overline{f}^{\mu\nu} \mathscr{l}_{\mu} \mathscr{l}_{\nu} \,, \quad &\mathscr{k}^{\mu} \,^{(f)}\nabla_{\mu} \mathscr{k}_{\nu} = 0 \, \quad \mathscr{l}^{\mu} \,^{(f)}\nabla_{\mu} \mathscr{l}_{\nu} &= 0\, . 
\end{align*}
The null vectors of each metric also follow orthogonality relations  given by
\begin{align*}
    k^{\mu}l_{\mu}=0= \mathscr{k}^{\mu}\mathscr{l_{\mu}}\,.
\end{align*}
The inverse metric of the double Kerr-Schild ansatz (\ref{ansatz_double ks bg}) can be written as:
\begin{align}
    \nonumber g^{\mu \nu} =&  \overline{g}^{\mu \nu} - \kappa_{g} \phi_{g} k^{\mu} k^{\nu} - \kappa_{g} \psi_{g} l^{\mu} l^{\nu} \,,\\
    \quad f^{\mu \nu} = & C^{-2}\left(\overline{f}^{*\,\mu \nu} - \kappa_{f} \phi_{f} \mathscr{k}^{\mu} \mathscr{k}^{\nu} - \kappa_{f} \psi_{f} \mathscr{l}^{\mu} \mathscr{l}^{\nu}\right) \,, 
\end{align}
which is an important property of this type of metrics and exploits the fact that $k_{\mu}$, $l_{\mu}$ and $\mathscr{k}_{\mu}$, $\mathscr{l}_{\mu}$ are null.  As we shall see later, we will restrict to the case in (\ref{ansatz_double ks bg}), where
$\overline{f}_{\mu\nu}=C^{2}\overline{g}_{\mu\nu}$ and the null vectors are the same for the two metrics, i.e., $\mathscr{k}_{\mu}=k_{\mu}$ and $\mathscr{l}_{\mu}=l_{\mu}$. This simplifies the interaction terms between the metrics, as it allows a closed form for the interaction matrix, as 
was first noted in \cite{Ayon-Beato:2015qtt}. For now we can consider the general case for the ansatz (\ref{ansatz_double ks bg}), where the vectors $k_{\mu}$, $l_{\mu}$ and $\mathscr{k}_{\mu},\mathscr{l}_{\mu} $ and the background metrics $\overline{g}_{\mu\nu}$ and $\overline{f}_{\mu\nu}$ are unrestricted.

If we consider two vacuum spacetimes $(\widetilde{S}_g,\widetilde{g}_{\mu \nu})$ and $(\overline{S}_g, \overline{g}_{\mu \nu})$ in GR related by a Kerr-Schild transformation, the null vector $k_{\mu}$ that appears in the transformation is geodetic with respect to both metrics  \cite{Stephani:2003tm}. For vacuum spacetimes in bigravity and generalized (single) Kerr-Schild ansatz, we can extend this result by imposing that the contraction of the tensors $V_{\mu\nu}$ and $\mathcal{V}_{\mu\nu}$ with the null vector of their respective metric is zero. Then, for vacuum spacetimes $S_{g}$, $\overline{S}_{g}$ and $S_{f}$, $\overline{S}_{f}$ in bigravity related by a Kerr-Schild transformation, the contraction of $V_{\mu\nu}$ and $\mathcal{V}_{\mu\nu}$ with the null vectors implies that the null vectors are geodetic with respect to both their corresponding full and background metrics. Correspondingly, for a double generalized Kerr-Schild ansatz in bigravity, we have that, if 
\begin{align*}
    V_{\mu\nu} \,k^{\mu}k^{\nu} = 0 = V_{\mu\nu} \,l^{\mu}l^{\nu} \, , \\
    \mathcal{V}_{\mu\nu}\, \mathscr{k}^{\mu}\mathscr{k}^{\nu} = 0 = \mathcal{V}_{\mu\nu}\, \mathscr{l}^{\mu}\mathscr{l}^{\nu}, 
\end{align*}
holds, the null vectors $k_{\mu}$, $l_{\mu}$ and $\mathscr{k}_{\mu}$, $\mathscr{l}_{\mu}$ are geodesic with respect their full and background metrics, i.e., 
\begin{align}
\label{geod}
    k^{\mu} \,^{(g)} {\nabla}_{\mu} k_{\nu} = &0 =k^{\mu} \,^{(g)}\overline{\nabla}_{\mu} k_{\nu} \,, \qquad  l^{\mu} \,^{(g)}{\nabla}_{\mu} l_{\nu} = 0 = l^{\mu} \,^{(g)}\overline{\nabla}_{\mu} l_{\nu} \,, \\
    \mathscr{k}^{\mu} \,^{(f)} {\nabla}_{\mu} \mathscr{k}_{\nu} = &0 = \mathscr{k}^{\mu} \,^{(f)}\overline{\nabla}_{\mu} \mathscr{k}_{\nu} \,, \qquad  \mathscr{l}^{\mu} \,^{(f)}{\nabla}_{\mu} \mathscr{l}_{\nu} = 0 = \mathscr{l}^{\mu} \,^{(f)}\overline{\nabla}_{\mu} \mathscr{l}_{\nu} \, .
\end{align}
We raise and lower the indices of the null vectors as follows:
\begin{align}
    \nonumber k^{\mu}=\overline{g}^{\mu\alpha}k_{\alpha} \, , \quad l^{\mu}={\overline{g}}^{\,\mu\alpha}l_{\alpha} \, ,\qquad \mathscr{k}^{\mu}=\overline{f}^{\mu\alpha}\mathscr{k}_{\alpha} \, , \quad \mathscr{l}^{\mu}={\overline{f}}^{\,\mu\alpha}\mathscr{l}_{\alpha} \, .
\end{align}
We also have that similar relations fulfilled for the background and full metrics, namely $\,^{(g)}\overline{\nabla}^{\mu}=\overline{g}^{\mu\alpha}\,^{(g)}\overline{\nabla}_{\alpha}, \, \,^{(f)}\overline{\nabla}^{\mu}=\overline{f}^{\,\mu\alpha}\,^{(f)}\overline{\nabla}_{\alpha}$ and $\,^{(g)}\nabla^{\mu}=g^{\mu\alpha}\,^{(g)}\nabla_{\alpha}, \, \,^{(f)}\nabla^{\mu}=f^{\,\mu\alpha}\,^{(f)}\nabla_{\alpha}$.

In the Kerr-Schild ansatz, the Ricci tensor with mixed indices $R^{\mu}{}_{\nu}$ is linear in the perturbation of the background metric. In the case of the double Kerr-Schild ansatz this no longer holds, but it can be linearized in some coordinates. Then, for the double Kerr-Schild ansatz and the $g$-metric, for example, the Ricci tensor has a non-linear contribution:
\begin{align*}
   R^{\mu}{}_{\nu}&=\overline{R}^{\mu}{}_{\nu}-\kappa_{g}\, h^{\mu\lambda} \overline{R}_{\lambda \nu}+\frac{\kappa_{g}}{2}\,^{(g)}\overline{\nabla}_{\lambda}\left[ \,^{(g)}\overline{\nabla}^{\mu}h^{\lambda}{}_{\nu}+\,^{(g)}\overline{\nabla}_{\nu}h^{\mu\lambda}-\,^{(g)}\overline{\nabla}^{\lambda}h^{\mu}{}_{\nu}\right] \, +R^{\mu}{}_{\nu,\text{NL}}
\end{align*}
where
\begin{align*}
    R^{\mu}{}_{\nu,\text{NL}}=&-\frac{\kappa^{2}}{2}\bigg[\frac{1}{2}\nabla^{\mu}h_{\phi\,\delta}^{\rho}\nabla_{\nu}h_{\psi\,\rho}^{\delta}+h_{\psi}^{\mu\delta}\nabla_{\rho}\nabla_{\nu}h_{\phi\,\delta}^{\rho} \\ &+\nabla_{\rho}\left(h_{\psi}^{\rho\delta}\nabla_{\delta}h_{\phi\,\nu}^{\mu}+2h_{\psi}^{\rho\delta}\nabla_{(\nu}h_{\phi\,\delta}^{\mu)}-2h_{\psi}^{\mu\delta}\nabla^{[\rho}h_{\phi\,\delta]\nu}\right)\bigg]+(k\leftrightarrow l)\, .
\end{align*}
Even though generically we have this non-linear contribution, we can choose some coordinates where it vanishes and the Ricci tensor linearizes. Then, we consider some coordinates where the Ricci tensor for both metrics linearizes in the total perturbation and, in such case, the Ricci tensors with mixed components for the generalized double Kerr-Schild ansatz (\ref{ansatz_double ks bg}) in bigravity takes the form:
\begin{align}
\label{Riccis_bigrav_ads}
    \nonumber R^{\mu}{}_{\nu}&=\overline{R}^{\mu}{}_{\nu}-\kappa_{g}\, h^{\mu\lambda} \overline{R}_{\lambda \nu}+\frac{\kappa_{g}}{2}\,^{(g)}\overline{\nabla}_{\lambda}\left[ \,^{(g)}\overline{\nabla}^{\mu}h^{\lambda}{}_{\nu}+\,^{(g)}\overline{\nabla}_{\nu}h^{\mu\lambda}-\,^{(g)}\overline{\nabla}^{\lambda}h^{\mu}{}_{\nu}\right] \, , \\
    \mathcal{R}^{\mu}{}_{\nu}&=\overline{\mathcal{R}}^{\mu}{}_{\nu}-C^{2}\kappa_{f} \, \mathscr{h}^{\mu\lambda} \overline{\mathcal{R}}_{\lambda\nu}+\frac{C^{2}\kappa_{f}}{2}\,^{(f)}\overline{\nabla}_{\lambda}\left[ \,^{(f)}\overline{\nabla}^{\mu}\mathscr{h}^{\lambda}{}_{\nu}+\,^{(f)}\overline{\nabla}_{\nu}\mathscr{h}^{\mu\lambda}-\,^{(f)}\overline{\nabla}^{\lambda}\mathscr{h}^{\mu}{}_{\nu}\right] \, ,
\end{align}
Consequently, we can write the equations of motion for the ansatz in (\ref{ansatz_double ks bg}) as:
\begin{align}
\label{eom_big_mat_h}
    \nonumber \kappa_{g}\,^{(g)}\overline{\mathcal{E}}(h^{\mu}{}_{\nu}) - \kappa_{g}\,h^{\mu\lambda} \overline{R}_{\lambda \nu}+ \overline{R}^{\mu}{}_{\nu} - \check{Q}^{\mu}{}_{\nu} \, = \, \frac{{\kappa_{g}}^{2}}{2} {\check{T}_{M}{}}^{\mu}{}_{\nu} \, , \\
    C^{2}\kappa_{f}\,^{(f)}\overline{\mathcal{E}}(\mathscr{h}^{\mu}{}_{\nu}) -C^{2} \kappa_{f} \, \mathscr{h}^{\mu\lambda} \overline{\mathcal{R}}_{\lambda\nu}+ \overline{\mathcal{R}}^{\mu}{}_{\nu} - \check{\mathcal{Q}}^{\mu}{}_{\nu}=  \frac{{\kappa_{f}}^{2}}{2} {\check{\mathcal{T}_M}}^{\mu}{}_{\nu}\, ,
\end{align}
with the total perturbations corresponding to a double Kerr-Schild form. We have used the Lichnerowicz operator in the form
\begin{align}
    \nonumber \,^{(g)}\overline{\mathcal{E}}(h^{\mu}{}_{\nu}) \equiv  \frac{1}{2}\,^{(g)}\overline{\nabla}_{\lambda}\left[ \,^{(g)}\overline{\nabla}^{\mu}h^{\lambda}{}_{\nu}+\,^{(g)}\overline{\nabla}_{\nu}h^{\mu\lambda}-\,^{(g)}\overline{\nabla}^{\lambda}h^{\mu}{}_{\nu}\right] \, , \\
    \nonumber \,^{(f)}\overline{\mathcal{E}}(\mathscr{h}^{\mu}{}_{\nu}) \equiv \frac{1}{2}\,^{(f)}\overline{\nabla}_{\lambda}\left[ \,^{(f)}\overline{\nabla}^{\mu}\mathscr{h}^{\lambda}{}_{\nu}+\,^{(f)}\overline{\nabla}_{\nu}\mathscr{h}^{\mu\lambda}-\,^{(f)}\overline{\nabla}^{\lambda}\mathscr{h}^{\mu}{}_{\nu}\right]\,.
\end{align}
The analysis for the double Kerr-Schild ansatz can be generalized to a multi-Kerr-Schild ansatz, given the linearization property of the Ricci tensor. 

After writing the result in (\ref{eom_big_mat_h}), we will work on computing the interaction tensors $Q_{\mu}{}_{\nu}$ and $\mathcal{Q}_{\mu}{}_{\nu}$ explicitly for the double Kerr-Schild ansatz in bigravity.

\subsection{Interaction between the metrics}
At this point we can consider the interaction between the two metrics, accounted by $Q^{\mu}{}_{\nu}$ and $\mathcal{Q}^{\mu}{}_{\nu}$ in (\ref{eom_big_mat_h}) and encoded in the interaction matrix $\gamma^{\mu}{}_{\nu}$ in (\ref{int_gamma}). In the same spirit of the ansatz (\ref{ansatz_1}), we will consider in this work the following double Kerr-Schild ansatz in bigravity:
\begin{align}
\label{ansatz_double ks bg2}
    \nonumber g_{\mu\nu} &= \overline{g}_{\mu\nu}+\kappa_{g}\,h_{\phi\,\mu\nu}+\kappa_{g}\, h_{\psi\,\mu\nu} \, ;  \quad &h_{\phi\,\mu\nu} = \phi_{g}k_{\mu}k_{\nu} \, , \quad  h_{\psi\, \mu\nu} &= \psi_{g}l_{\mu}l_{\nu} \\
    f_{\mu\nu}\, &=  \, C^{2}(\overline{g}_{\mu\nu}+\kappa_{f}\mathscr{h}_{\phi\, \mu\nu} + \kappa_{f}\mathscr{h}_{\psi\, \mu\nu}) \, ; \quad &\mathscr{h}_{\phi\, \mu\nu} = \phi_{f} k_{\mu}k_{\nu} \, , \quad \mathscr{h}_{\psi\, \mu\nu} &= \psi_{f} l_{\mu}l_{\nu}  \, , 
\end{align}
and we obtain that the powers of the interaction matrix has the form
\begin{align}
    \nonumber \left(\gamma^{n}\right)^{\mu}{ }_{\nu}=C^{n}\left[\delta^{\mu}{ }_{\nu}-\frac{n}{2}\left(\kappa_{g}h^{\mu}{}_{\nu}-C^{2}\kappa_{f}\mathscr{h}^{\mu}{}_{\nu}\right) \right] \, ,
\end{align}
where we have the full perturbations as in (\ref{full_pert}) for the double KS ansatz. This form is analogue to the one obtained in  \cite{Ayon-Beato:2015qtt} for the generalized single Kerr-Schild ansatz. We obtain similar results as \cite{Garcia-Compean:2024uie}, where the potential $\mathcal{U}[g,f]$ and $\tau^{\mu}{}_{\nu}$ have the form:
\begin{align}
    \nonumber \mathcal{U}[g,f] =-(P_{1}+C P_{2}) \, , \quad \tau^{\mu}{}_{\nu} =-C\left[P_{2}\, \delta^{\mu}{}_{\nu} + P_{0} \left(\kappa_{g}h^{\mu}{}_{\nu}-C^{2}\kappa_{f}\mathscr{h}^{\mu}{}_{\nu}\right)\right] \, . 
\end{align}
We have defined the following quantities in terms of the coupling constants $b_{k}$ that appear in the action (\ref{action_big}):
\begin{align}
    \nonumber P_{0} &\equiv \frac{1}{2}\left(b_{1}+2Cb_{2}+C^{2}b_{3}\right) \, , \\
    \nonumber P_{1} &\equiv -\left(b_{0}+3 C b_{1}+3 C^{2} b_{2}+C^{3} b_{3}\right) \, , \\
    \nonumber P_{2} &\equiv -\left(b_{1}+3 C b_{2}+3 C^{2} b_{3}+C^{3} b_{4}\right)\, .
\end{align}
In this case the interaction tensors can be expressed as:
\begin{align}
\label{int_tensors_ads_2}
    \nonumber & Q^{\mu}{}_{\nu} = \kappa_{g}\left[B_{0} \delta^{\mu}{}_{\nu} - B_{1} \left(\kappa_{g}h^{\mu}{}_{\nu}-C^{2}\kappa_{f}\mathscr{h}^{\mu}{}_{\nu}\right)\right]  \, , \quad Q = 4\, \kappa_{g}\, B_{0} \, , \\
    & \mathcal{Q}^{\mu}{}_{\nu} = C^{2}\kappa_{f}\left[\,\mathcal{B}_{0}\delta^{\mu}{}_{\nu} + \mathcal{B}_{1} \left(\kappa_{g}h^{\mu}{}_{\nu}-C^{2}\kappa_{f}\mathscr{h}^{\mu}{}_{\nu}\right)\right]\, , \quad \mathcal{Q}=4\,C^{2}\kappa_{f}\,\mathcal{B}_{0}\, ,
\end{align}
where we have defined:
\begin{align}
    \nonumber B_{0} = \frac{m^{2} \kappa_{g}}{\kappa^{2}}(P_{1}) \, , \quad \mathcal{B}_{0} = \frac{m^{2} \kappa_{f}}{\kappa^{2}}\left(\frac{P_{2}}{C^{5}}\right) \, , \quad B_{1} = \frac{m^{2} \kappa_{g}}{\kappa^{2}}(C P_{0}) \, , \quad  \mathcal{B}_{1} = \frac{m^{2} \kappa_{f}}{\kappa^{2}}\left(\frac{P_{0}}{C^{5}}\right) \, . 
\end{align}
Then, the trace-reversed equations of bigravity coupled to matter for the ansatz in (\ref{ansatz_double ks bg}) can be written as:
\begin{align}
\label{eom_matter}
    \nonumber \,^{(g)}\overline{\mathcal{E}}(h^{\mu}{}_{\nu}) - h^{\mu\lambda} \overline{R}_{\lambda \nu}+\frac{\overline{R}^{\mu}_{\nu}}{\kappa_{g}} = &\,  - B_{0}\,\delta^{\mu}{}_{\nu} - B_{1} \left(\kappa_{g}h^{\mu}{}_{\nu}-C^{2}\kappa_{f}\mathscr{h}^{\mu}{}_{\nu}\right) + \frac{\kappa_{g}}{2} {\check{T}_{M}{}}^{\mu}{}_{\nu}\, , \\
    \,^{(f)}\overline{\mathcal{E}}(\mathscr{h}^{\mu}{}_{\nu}) - \mathscr{h}^{\mu\lambda} \overline{\mathcal{R}}_{\lambda\nu}+\frac{\overline{\mathcal{R}}^{\mu}{}_{\nu}}{C^{2} \kappa_{f}} = & \, - \mathcal{B}_{0}\, \delta^{\mu}{}_{\nu} + \mathcal{B}_{1} \left(\kappa_{g}h^{\mu}{}_{\nu}-C^{2}\kappa_{f}\mathscr{h}^{\mu}{}_{\nu}\right)+\frac{\kappa_{f}}{2} {\check{\mathcal{T}_M}}^{\mu}{}_{\nu}\, , 
\end{align}
where
\begin{align*}
    & h_{\mu\nu}=\phi_{g}k_{\mu}k_{\nu}+\psi_{g}l_{\mu}l_{\nu} \, , \quad \mathscr{h}_{\mu\nu}=\phi_{f}k_{\mu}k_{\nu}+\psi_{f}l_{\mu}l_{\nu}\, .
\end{align*}
These are the equations of motion of bigravity in terms of the spin-2 full perturbations to the background metrics coupled to matter.

In order to study the equations (\ref{eom_matter}) in the context of the double copy, the work in \cite{Carrillo-Gonzalez:2017iyj} provides a great framework, as they present a procedure in GR to obtain the single and zeroth copy equations for some generalized Kerr-Schild stationary and time-dependent solutions in maximally symmetric backgrounds by contracting the field equations with Killing vectors. In \cite{Garcia-Compean:2024uie}, we obtained a similar structure for bigravity and particularly for the ansatz in (\ref{ansatz_1}). In what follows, we will present a formalism for the generalized double Kerr-Schild ansatz in bigravity coupled to matter in maximally symmetric backgrounds.

\subsection{Double Kerr-Schild in maximally symmetric backgrounds}
In order to write the single and zeroth copy equations of bigravity for the ansatz (\ref{ansatz_double ks bg2}), we will make some simplifications and provide some definitions. We will focus on solutions where $B_{0}$ and $\mathcal{B}_{0}$ are fixed by the diagonal contributions to the equations. This restriction holds for the stationary and time-dependent solutions we will present and can be written as:
\begin{align}
    \label{diag_constraint}
    \frac{\overline{R}^{\mu}{}_{\nu}}{\kappa_{g}} + B_{0}\,\delta^{\mu}{}_{\nu}\, = 0 \, ,  \quad  \frac{ \overline{\mathcal{R}}^{\mu}{}_{\nu}}{C^{2}\,\kappa_{f}} + \mathcal{B}_{0}\, \delta^{\mu}{}_{\nu}\, = 0 \, .
\end{align}
We are fixing the values of the $b_{k}$ in $P_{0},P_{1}$ and $P_{2}$, and this will simplify the expressions we will work on. On the other side, for maximally symmetric backgrounds, analog to GR, the  background Ricci tensors satisfy:
\begin{align}
    \nonumber \overline{R}_{\mu\nu}=\Lambda_{g} \overline{g}_{\mu\nu} \, , \quad \overline{R}=4 \Lambda_{g}\,  \, \qquad
    \nonumber \overline{\mathcal{R}}_{\mu\nu}=\Lambda_{f} \overline{f}_{\mu\nu} \, , \quad \overline{\mathcal{R}}=4 \Lambda_{f}\, ,
\end{align}
where $\Lambda_{g}$ and $\Lambda_{f}$ are the cosmological constants for the background metrics $\overline{g}_{\mu\nu}$ and $\overline{f}_{\mu\nu}$. For the ansatz in (\ref{ansatz_double ks bg2}) we obtain that $\overline{R}=C^{2}\overline{\mathcal{R}}$. 

Another ingredient needed for analyzing the ansatz in the double copy context is the definition of the fields: 
\begin{align} 
\label{def_gauge_fields}
    \nonumber A_{\phi\,\mu} \equiv \, & \phi_{g} k_{\mu} \, , \quad A_{\phi}^{\mu}\equiv \overline{g}^{\mu \alpha} A_{\phi\,\alpha} \, , \quad  F_{\phi}^{\mu\nu}=\overline{\nabla}^{\mu}A_{\phi}^{\nu}-\overline{\nabla}^{\nu}A_{\phi}^{\mu} \, , \\
    \mathcal{A}_{\phi\,\mu}= \, & \phi_{f}k_{\mu} \, , \quad  \mathcal{A}_{\phi}^{\mu}\equiv \overline{f}^{\mu \alpha} \mathcal{A}_{\phi\,\alpha} \, , \quad \mathcal{F}_{\phi}^{\mu\nu}=\,^{(f)}\overline{\nabla}^{\mu}\mathcal{A}_{\phi}^{\nu}-\,^{(f)}\overline{\nabla}^{\nu}\mathcal{A}_{\phi}^{\mu} \, .
\end{align}
Moreover we define the following quantities:
\begin{align}
    \nonumber & W^{\mu}_{\phi\,\nu}\equiv X^{\mu}_{\phi\,\nu}+Y^{\mu}{}_{\nu} \, , \quad \mathcal{W}^{\mu}_{\phi\,\nu}\equiv\mathcal{X}^{\mu}_{\phi\,\nu}+\mathcal{Y}^{\mu}_{\phi\,\nu} \,, \\
    \nonumber & {X}^{\mu}_{\phi\,\nu} \equiv-\overline{\nabla}_{\nu}\left(k^{\mu}\overline{\nabla}_\lambda A_{\phi}^{\lambda}\right) \, , \quad Y^{\mu}_{\phi\,\nu} \equiv F_{\phi}^{\rho\mu}\,\overline{\nabla}_{\rho}k_{\nu}-\overline{\nabla}_{\rho}\left(A_{\phi}^{\rho}\overline{\nabla}^{\mu}k_{\nu}-A_{\phi}^{\mu}\overline{\nabla}_{\rho}k_{\nu}\right) \, , \\
    \nonumber & \mathcal{X}^{\mu}_{\phi\,\nu} \equiv-\overline{\nabla}_\nu\left(\mathscr{k}^\mu\overline{\nabla}_\lambda \mathcal{A}_{\phi}^{\lambda}\right) \, , \quad \mathcal{Y}^{\mu}_{\phi\,\nu} \equiv \mathcal{F}_{\phi}^{\rho\mu}\,\overline{\nabla}_{\rho}\mathscr{k}_{\nu}-\overline{\nabla}_{\rho}\left(\mathcal{A}_{\phi}^{\rho}\,^{(f)}\overline{\nabla}^{\mu}\mathscr{k}_{\nu}-\mathcal{A}_{\phi}^{\mu}\overline{\nabla}_{\rho}\mathscr{k}_{\nu}\right) \, \\
    \nonumber & Z_{\phi}^{\mu} \equiv \,(K_{\lambda}k^{\lambda})\left[\overline{\nabla}_{\sigma}k^{\mu} \, \overline{\nabla}^{\sigma}\phi_{g}+\overline\nabla_{\sigma}(\phi_{g}\overline\nabla^{\sigma}k^{\mu}-\phi_{g}\overline\nabla^{\mu}k^{\sigma}-k^{\sigma}\overline\nabla^{\mu}\phi_{g})\right] \, , \\
    \nonumber  &\mathcal{Z}_{\phi}^{\mu} \equiv \, (\mathcal{K}_{\lambda}\mathscr{k}^{\lambda})\left[\overline{\nabla}_{\sigma}\mathscr{k}^{\mu} \, \,^{(f)}\overline{\nabla}^{\sigma}\phi_{f}+\overline{\nabla}_{\sigma}\left(\phi_{f}\,^{(f)}\overline{\nabla}^{\sigma}\mathscr{k}^{\mu}-\phi_{f}\,^{(f)}\overline\nabla^{\mu}\mathscr{k}^{\sigma}-l^{\sigma}\,^{(f)}\overline\nabla^{\mu}\phi_{f}\right)\right] \, , \\
\end{align}
and the analogue quantities with the functions $\psi_{g}$, $\psi_{f}$ and the null vectors $l_{\mu}$, $\mathscr{l}_{\mu}$, i.e., $A_{\psi\,\mu}$, $F_{\psi}^{\mu\nu}$, ${X}^{\mu}_{\psi\,\nu}$, $Y^{\mu}_{\psi\,\nu}$, $Z_{\psi}^{\mu}$ and $\mathcal{A}_{\psi\,\mu}$, $\mathcal{F}_{\psi}^{\mu\nu}$, $\mathcal{X}^{\mu}_{\psi\,\nu}$, $\mathcal{Y}^{\mu}_{\psi\,\nu}$, $\mathcal{Z}_{\psi}^{\mu}$. These quantities are defined as in \cite{Carrillo-Gonzalez:2017iyj} for general relativity.
We also have: 
\begin{align}
    \nonumber F_{\mu\nu}&\equiv F_{\phi\,\mu\nu}+F_{\psi\,\mu\nu}\,, \quad  \mathcal{F}_{\mu\nu}\equiv \mathcal{F}_{\phi\,\mu\nu}+\mathcal{F}_{\psi\,\mu\nu} \, ,\\
    \nonumber X^{\mu}{}_{\nu} &\equiv X^{\mu}_{\phi\,\nu}+X^{\mu}_{\psi\,\nu} \, , \quad Y^{\mu}{}_{\nu}\equiv Y^{\mu}_{\phi\,\nu}+Y^{\mu}_{\psi\,\nu} \, , \quad   Z^{\mu}{}_{\nu} \equiv Z^{\mu}_{\phi\,\nu}+Z^{\mu}_{\psi\,\nu} \, ,\\
    \mathcal{X}^{\mu}{}_{\nu}& \equiv \mathcal{X}^{\mu}_{\phi\,\nu}+\mathcal{X}^{\mu}_{\psi\,\nu} \, , \quad \mathcal{Y}^{\mu}{}_{\nu}\equiv \mathcal{Y}^{\mu}_{\phi\,\nu}+\mathcal{Y}^{\mu}_{\psi\,\nu} \,  , \quad \mathcal{Z}^{\mu}{}_{\nu} \equiv \mathcal{Z}^{\mu}_{\phi\,\nu}+\mathcal{Z}^{\mu}_{\psi\,\nu}\, . 
\end{align}
We also have
\begin{align}
    \nonumber & W^{\mu}{}_{\nu}\equiv X^{\mu}{}_{\nu}+Y^{\mu}{}_{\nu} \, , \quad \mathcal{W}^{\mu}{}_{\nu}\equiv\mathcal{X}^{\mu}{}_{\nu}+\mathcal{Y}^{\mu}{}_{\nu}\\   
    \nonumber & \alpha \equiv \, K^{\rho}k_{\rho} = K^{\alpha}l_{\alpha} \, , \quad  \beta \equiv \, \mathcal{K^{\rho}}\mathscr{k}_{\rho} = \mathcal{K}^{\alpha} \mathscr{l}_{\alpha}  \, ,
\end{align}
and we note that the contraction of Killing vector with both null vectors of the metric is the same. With these definitions and the constraint (\ref{diag_constraint}), for maximally symmetric backgrounds, the bigravity equations (\ref{eom_big}) for the ansatz in (\ref{ansatz_double ks bg2}) are written as
\begin{align}
    \label{eom_dc_1}
    \nonumber \,^{(g)}\overline{\mathcal{E}}(h^{\mu}{}_{\nu}) -\frac{\overline{R}}{4} h^{\mu}{}_{\nu} = &\, - B_{1} \left(\kappa_{g}h^{\mu}{}_{\nu}-C^{2}\kappa_{f}\mathscr{h}^{\mu}{}_{\nu}\right) + \frac{\kappa_{g}}{2} {\check{T}_{M}{}}^{\mu}{}_{\nu}\, , \\
    \,^{(f)}\overline{\mathcal{E}}(\mathscr{h}^{\mu}{}_{\nu}) -\frac{\overline{\mathcal{R}}}{4} \mathscr{h}^{\mu}{}_{\nu} = & \, \mathcal{B}_{1} \left(\kappa_{g}h^{\mu}{}_{\nu}-C^{2}\kappa_{f}\mathscr{h}^{\mu}{}_{\nu}\right)+\frac{\kappa_{f}}{2} {\check{\mathcal{T}_M}}^{\mu}{}_{\nu}\, .
\end{align}
These equations can be rewritten and contracted with Killing vectors and, by making the change of the coupling constant as proposed in the KS double copy procedure in bigravity, we obtain the single copy equations:
\begin{align}
    \label{eom_sc_1}
    \nonumber \overline{\nabla}_{\nu}F^{\nu\mu}+\frac{\overline{R}}{6}A^{\mu}+\frac{1}{\alpha}K^{\nu} W^{\mu}{}_{\nu}\, = & \, 2\,B_{1} \left(\kappa_{g}A^{\mu}-C^{2}\kappa_{f}\mathcal{A}^{\mu}\right) + g_{g}\, {J_{M}{}}^{\mu}\, ,  \\
    \overline{\nabla}_{\nu}\mathcal{F}^{\nu\mu} + \frac{\overline{\mathcal{R}}}{6}\mathcal{A}^{\mu}+\frac{1}{\beta}\mathcal{K}^{\nu}\mathcal{W}^{\mu}{}_{\nu} \, = &\, -2\, \mathcal{B}_{1} \left(\kappa_{g}A^{\mu}-C^{2}\kappa_{f}\mathcal{A}^{\mu}\right)+g_{f}\, {\mathcal{J}_{M}{}}^{\mu} \, .
\end{align}
Proceeding further we can obtain the zeroth copy equations:
\begin{align}
\label{eom_zc_1}
    \nonumber \overline{\nabla}^{2}\Upsilon_{g}+\frac{\overline{R}}{6}\Upsilon_{g}+\frac{1}{\alpha^2}K_{\mu}\left(K^{\nu}W^{\mu}{}_{\nu}+Z^{\mu}\right)\, = \, & 2 B_{1} \left(\kappa_{g}\Upsilon_{g}-\kappa_{f}\Upsilon_{f}\right)+y_{g}\,j_{M} \, , \\
    \,^{(f)}\overline{\nabla}^{2}\Upsilon_{f} +\frac{\overline{\mathcal{R}}}{6}\Upsilon_{f}+\frac{1}{\beta^2}\mathcal{K}_{\mu}\left(\mathcal{K}^{\nu}\mathcal{W}^{\mu}{}_{\nu}+\mathcal{Z}^{\mu}\right) \, = \, & - 2\, C^{2}\,\mathcal{B}_{1} \left(\kappa_{g}\Upsilon_{g}-\kappa_{f}\Upsilon_{f}\right)+y_{f}\,\mathcal{j}_{M} \, ,
\end{align}
where we have also changed the coupling constants. The full fields are:
\begin{align}
    \label{full_fields}
    \nonumber & h_{\mu\nu}=h_{\phi\,\mu\nu}+ h_{\psi\,\mu\nu}=\phi_{g}k_{\mu}k_{\nu}+\psi_{g}l_{\mu}l_{\nu} \, , \\
    \nonumber & \mathscr{h}_{\mu\nu}=\mathscr{h}_{\phi\,\mu\nu}+ \mathscr{h}_{\psi\,\mu\nu}=\phi_{f}k_{\mu}k_{\nu}+\psi_{f}l_{\mu}l_{\nu}\, ,\\ 
    \nonumber & A_{\mu}=A_{\phi\,\mu}+A_{\psi\,\mu}\, = \phi_{g}k_{\mu}+\psi_{g}l_{\mu},\\
    \nonumber &\mathcal{A}_{\mu}=\mathcal{A}_{\phi\,\mu}+ \mathcal{A}_{\psi\,\mu}=\phi_{f}k_{\mu}+ \psi l_{\mu}\, ,\\
    &\Upsilon_{g}=\phi_{g}+\psi_{g}\, , \quad  \Upsilon_{f}=\phi_{f}+\psi_{f}  \, .   
\end{align}
We have used the trace-reversed energy-momentum tensors ${\check{T}_{M}{}}^{\mu}{}_{\nu}$ and ${\check{\mathcal{T}}_{M}{}}^{\mu}{}_{\nu}$ to define the single and zeroth copy sources as:
\begin{align}
    \nonumber {J_{M}}^{\mu} &\equiv\frac{- 2 K^{\nu}}{\alpha}{\check{T}_{M}{}}^{\mu}{}_{\nu} \, ,\quad {\mathcal{J}_{M}}^{\mu}\equiv\frac{- 2 \mathcal{K}^{\nu}}{C^{2} \beta}{\check{\mathcal{T}}_{M}{}}^{\mu}{}_{\nu}\,, \qquad
    \nonumber j_{M} &\equiv \frac{1}{\alpha}K_{\mu}{J_{M}}^{\mu} \, , \quad {\mathscr{j}_{M}}\equiv \frac{1}{\beta}\mathcal{K}_{\mu}{\mathcal{J}_{M}}^{\mu}\, .
\end{align}
We have obtained these previous equations by rewriting and contracting the equations (\ref{eom_matter}) with Killing vectors $K^{\mu}$ for the metric $g_{\mu\nu}$ and $\mathcal{K}^{\mu}$ for the metric $f_{\mu\nu}$ as shown in \cite{Garcia-Compean:2024uie}. In the examples we will present, we use  timelike Killing vectors for stationary solutions and null Killing vectors for time-dependent solutions. 

By exploiting again the relations between covariant derivatives, we can rewrite the equations of motion using only one differential operator, $\,^{(g)}\overline{\nabla}_{\mu}$ by redefining the fields
\begin{align}
\label{redef_ads_2}
    \nonumber & h^{+\,\mu}{}_{\nu}\equiv -\frac{\kappa_{g}\kappa_{f}}{\kappa^{2}}\left(\kappa_{f}\,h^{\mu}{}_{\nu}+C^{2}\,\kappa_{g}\,\tilde{\mathscr{h}}^{\mu}{}_{\nu}\right)\, , \quad h^{-\,\mu}{}_{\nu}\equiv \kappa_{g}\, h^{\mu}{}_{\nu}-\kappa_{f}\,\tilde{\mathscr{h}}^{\mu}{}_{\nu}\, ,\\
    \nonumber & A^{+\, \mu} \equiv -\frac{\kappa_{g}\kappa_{f}}{\kappa^{2}}\left(\kappa_{f}\, A^{\mu}+C^{2}\, \kappa_{g}\,\tilde{\mathcal{A}^{\mu}}\right) \, , \quad  A^{-\, \mu} \equiv \kappa_{g}\,A^{\mu}- \kappa_{f}\,\tilde{\mathcal{A}^{\mu}} \, ,\\
    &\Upsilon^{+} \equiv -\frac{\kappa_{g}\kappa_{f}}{\kappa^{2}}\left(\kappa_{f} \, \Upsilon_{g}+C^{2}\, \kappa_{g} \, \Upsilon_{f}\right)\, , \quad  \Upsilon^{-} \equiv \kappa_{g}\,\Upsilon_{g}-\kappa_{f}\,  \Upsilon_{f}\,,  
\end{align}
 and by decoupling the equations of motion at the level of the double, single and zeroth copy. We have used the tensors 
 $\tilde{\mathscr{h}}^{\mu}{}_{\nu}$ and $\tilde{\mathcal{A}}^{\mu}$ given by:
 \begin{align}
    \nonumber \tilde{\mathscr{h}}^{\mu}{}_{\nu}\equiv \overline{g}^{\mu\alpha}\mathscr{h}_{\alpha\nu}= C^{2}\mathscr{h}^{\mu}{}_{\nu} \, , \quad  \tilde{\mathcal{A}}^{\mu}\equiv \overline{g}^{\mu\alpha}\mathcal{A}_{\alpha}= C^{2}\mathcal{A}^{\mu} \, , 
\end{align}
instead of $\mathscr{h}^{\mu}{}_{\nu}$ and $\mathcal{A}^{\mu}$, which may be interpreted as a translation of an object defined in the metric $f_{\mu\nu}$ to the metric $g_{\mu\nu}$. We can invert the relation (\ref{redef_ads_2}) by using the formula:
\begin{align}
    \label{redef_invs}
    \nonumber {\phi_{g}} =\frac{1}{\kappa_{g} \left(C^{2} \kappa_{g}^{2}+{\kappa_{f}}^{2}\right)}\bigg(C^{2}{\kappa_g}^{2}\,{\phi^{-}}-\kappa^{2}\, {\phi^{+}} \bigg)\, , \\
    {\phi_{f}} =-\frac{1}{\kappa_{f} \left(C^2 {\kappa_g}^{2}+{\kappa_{f}}^{2}\right)}\bigg({\kappa_f}^{2}\,{\phi^{-}}+\kappa^{2}\,{\phi^{+}}\bigg) \, , 
\end{align}
and analog expressions for the other fields. 

The formalism to which we have arrived in the equations in (\ref{eom_dc_1}), (\ref{eom_sc_1}) and (\ref{eom_zc_1}) for the double Kerr-Schild ansatz in (\ref{ansatz_double ks bg2}) reduces to the one presented in \cite{Garcia-Compean:2024uie} for a generalized Kerr-Schild ansatz in bigravity for the vacuum case, ${T_{M}}^{\mu}{}_{\nu}=0={\mathcal{T}_{M}}^{\mu}{\nu}$, when considering and $\psi_{g}=0=\psi_{f}$, so that $h_{\mu\nu}=h_{\phi\,\mu\nu}, \mathscr{h}_{\mu\nu}=\mathscr{h}_{\phi\,\mu\nu}$, and all the quantities defined with these, e.g, $X^{\mu}_{\psi\,\nu}$, $\mathcal{X}^{\mu}_{\psi\,\nu}$ equal to zero. Then, recover the equations of motion for the single Kerr-Schild proportional ansatz. For example, in the non-vacuum case, at the zeroth copy level, the equations (\ref{gf_evac_eom}) with $\psi_{g}=0=\psi_{f}$ imply:
\begin{align}
    \nonumber \overline{\nabla}^{2}\phi_{g}+\frac{\overline{R}}{6}\phi_{g}+\frac{1}{\alpha^2}K_{\mu}\left(K^{\nu}W^{\mu}_{\phi\,\nu}+Z^{\mu}_{\phi}\right)\, = \, & 2 B_{1} \left(\kappa_{g}\phi_{g}-\kappa_{f}\phi_{f}\right)+y_{g}\,j_{M} \, , \\
    \,^{(f)}\overline{\nabla}^{2}\phi_{f} +\frac{\overline{\mathcal{R}}}{6}\phi_{f}+\frac{1}{\beta^2}\mathcal{K}_{\mu}\left(\mathcal{K}^{\nu}\mathcal{W}^{\mu}_{\phi\,\nu}+\mathcal{Z}^{\mu}_{\phi}\right) \, = \, & - 2\, C^{2}\,\mathcal{B}_{1} \left(\kappa_{g}\phi_{g}-\kappa_{f}\phi_{f}\right)+y_{f}\,\mathcal{j}_{M} \, ,
\end{align}
which are the ones in \cite{Garcia-Compean:2024uie} with an additional matter term. The resulting equations will depend on the solution studied and can be simplified in specific coordinates.

\clearpage
\section{Single Kerr-Schild: AdS waves in bigravity coupled to matter} \label{AdS waves coupled to matter in BG}

We will examine first a single Kerr-Schild ansatz in bigravity. Let us consider AdS waves solutions in bigravity coupled to matter. In particular, the case of the Siklos-AdS waves. These solutions are written in a generalized single Kerr-Schild form, so that:
\begin{align*}
    \psi_{g}=0=\psi_{f}\, .
\end{align*}
Then, $h_{\psi\,\mu\nu}=0=\mathscr{h}_{\psi\,\mu\nu}$ and the Ricci tensor linearizes. Then, $h_{\mu\nu}=h_{\phi\,\mu\nu}$, $\mathscr{h}_{\mu\nu}=\mathscr{h}_{\phi\,\mu\nu}$ for the double, single and zeroth copy fields of both metrics, and we would also have something similar for the quantities we have defined, e.g., $X^{\mu}{}_{\nu}=X^{\mu}_{\phi\,\nu}$ and setting $X^{\mu}_{\psi\,\nu}=0$ for all these quantities defined with $\psi_{g}$ and $\psi_{f}$. 

For this case of AdS waves in bigravity, which is given in Poincaré coordinates, we have that
\begin{align}
\label{waves_constraints}
    X^{\mu}{}_{\nu} = 0 = \, Y^{\mu}{}_{\nu} \, , \quad \mathcal{X}^{\mu}{}_{\nu} =0= \mathcal{Y}^{\mu}{}_{\nu} \, , \quad \frac{1}{\alpha^2}K_{\mu}Z^{\mu} \, = \, \, - \frac{\overline{R}}{6}\phi_{g} \, , \quad \frac{1}{\beta^2}\mathcal{K}_{\mu}\mathcal{Z}^{\mu} \, = \, -\frac{\overline{\mathcal{R}}}{6}\phi_{f} \, , 
\end{align}
with no restriction on $P_{0}$. Equations in (\ref{waves_constraints}) hold for both the vacuum and the non-vacuum cases. In these cases, we have linear contributions to the interaction terms that depend on the perturbations. In contrast to stationary solutions, the conditions in (\ref{waves_constraints}) hold for the vacuum solutions studied in \cite{Garcia-Compean:2024uie} and also for the solutions coupled to matter that we will study here. Then, with these restrictions we have:
\begin{align}
    \label{time_dep_eom}
    \nonumber \,^{(g)}\overline{\mathcal{E}}(h^{\mu}{}_{\nu}) - \frac{\overline{R}}{4} \, h^{\mu}{}_{\nu} \,  = &\,  - \nonumber B_{1} \left(\kappa_{g}h^{\mu}{}_{\nu}-C^{2}\kappa_{f}\mathscr{h}^{\mu}{}_{\nu}\right) +\frac{\kappa_{g}}{2} {\check{T}_{M}{}}^{\mu}{}_{\nu}\, , \\
    \nonumber \,^{(f)}\overline{\mathcal{E}}(\mathscr{h}^{\mu}{}_{\nu}) -\,\frac{\overline{\mathcal{R}}}{4} \, \mathscr{h}^{\mu}{}_{\nu} \,\, = & \, \mathcal{B}_{1}\left(\kappa_{g}h^{\mu}{}_{\nu}-C^{2}\kappa_{f}\mathscr{h}^{\mu}{}_{\nu}\right)+\frac{\kappa_{f}}{2 \,C^{2}}{\check{\mathcal{T}}_{M}{}}^{\mu}{}_{\nu}\, ,\\
    \nonumber \overline{\nabla}_{\lambda}F^{\lambda\mu}+\frac{\overline{R}}{6}A^{\mu} \, = & \, 2 B_{1} \left(\kappa_{g}A^{\mu}-C^{2}\kappa_{f}\mathcal{A}^{\mu}\right) + g_{g} {J_{M}{}}^{\mu}\, , \\
    \nonumber \,^{(f)}\overline{\nabla}_{\lambda}\mathcal{F}^{\lambda\mu} +\frac{\overline{\mathcal{R}}}{6}\mathcal{A}^{\mu} \, = & \, -2\, \mathcal{B}_{1} \left(\kappa_{g}A^{\mu}-C^{2}\kappa_{f}\mathcal{A}^{\mu}\right)+g_{f} {\mathcal{J}_{M}{}}^{\mu} \, ,\\
        \nonumber \overline{\nabla}^{2}\phi_{g}\,  = \, &  2 B_{1} \left(\kappa_{g}\phi_{g}-\kappa_{f}\phi_{f}\right) +y_{g}j_{M} \, , \\
    \,^{(f)}\overline{\nabla}^{2}\phi_{f} \, = \, & - 2\,C^{2}\,\mathcal{B}_{1} \left(\kappa_{g}\phi_{g}-\kappa_{f}\phi_{f}\right)+y_{f} \mathcal{j}_{M} \, ,
\end{align}
where  $h_{\mu\nu}=h_{\phi\,\mu\nu}$, $\mathscr{h}_{\mu\nu}=\mathscr{h}_{\phi\,\mu\nu}$ and so on. We have also changed the gravitational constants to the respective coupling constants at each level.

As can be seen from (\ref{time_dep_eom}), the equations of motion for time-dependent solutions are coupled. By rewriting the equations using a single derivative operator $\,^{(g)}\overline{\nabla}_{\mu}$ and with aid of the redefinitions in (\ref{redef_ads_2}), we obtain that the decoupled fields "$+$" and "$-$" satisfy the following equations:
\begin{align}
\label{decoup_eqns}
    \nonumber  \,^{(g)}\overline{\mathcal{E}}(h^{+\,\mu}{}_{\nu})  +\frac{1}{2}{m^{+}_{D}}^{2} h^{+\,\mu}{}_{\nu} \, = & \, {\check{T}^{+}_{M}{}}^{\mu}{}_{\nu}  \, , \qquad {m^{+}_{D}}^{2} \equiv - \frac{\overline{R}}{2} \, , \\
    \nonumber \,^{(g)}\overline{\mathcal{E}}(h^{-\,\mu}{}_{\nu}) + \frac{1}{2} {m^{-}_{D}}^{2} \, h^{-\, \mu}{}_{\nu} \, = & \, {\check{T}^{-}_{M}{}}^{\mu}{}_{\nu} \, , \qquad {m^{-}_{D}}^{2} \equiv \widehat{m}^{2} - \frac{\overline{R}}{2} \, , \\
    \nonumber \overline{\nabla}_{\lambda}F^{+\,\lambda\mu} - {m^{+}_{S}}^{2} \, A^{+ \, \mu} = &\,  {J_{M}^{+}{}}^{\mu}  \, , \qquad {m^{+}_{S}}^{2} \equiv -\frac{\overline{R}}{6} \, ,\\
    \nonumber \overline{\nabla}_{\lambda}F^{-\,\lambda\mu} - {m^{-}_{S}}^{2}\,  A^{- \, \mu} \, = & \, \, {J_{M}^{-}{}}^{\mu} \, , \qquad {m^{-}_{S}}^{2} \equiv \widehat{m}^{2} -\frac{\overline{R}}{6}\, , \\
    \nonumber \overline{\nabla}^{2}\phi^{+}= &\, {j_{M}^{+}}  \, , \qquad {m^{+}_{Z}}^{2} \equiv 0  \, , \\ 
    \overline{\nabla}^{2}\phi^{-} - {m^{-}_{Z}}^{2} \, \phi^{-} \, = &\, \, {j_{M}^{-}} \, , \qquad {m^{-}_{Z}}^{2} \equiv \widehat{m}^{2} \, ,
\end{align}
where the decoupled trace-reversed energy-momentum tensors from the bigravity side are:
\begin{align}
    \nonumber {\check{T}_{M}^{+}{}}^{\mu}{}_{\nu} \equiv \frac{-\kappa_{g}\kappa_{f}}{\kappa^{2}}\left({\check{T}_{M}{}}^{\mu}{}_{\nu}+C^2 {\widetilde{\check{\mathcal{T}}}_{M}{}}^{\mu}{}_{\nu}\right) \, , \quad 
    {\check{T}_{M}^{-}{}}^{\mu}{}_{\nu} \equiv \frac{{\kappa_{g}}^{2}}{2} {\check{T}_{M}{}}^{\mu}{}_{\nu} - \frac{{\kappa_{f}}^{2}}{2}{\widetilde{\check{\mathcal{T}}}_{M}{}}^{\mu}{}_{\nu} \, .
\end{align}
They correspond to the effective energy-momentum tensor each of the decoupled fields perceives. The tilde in the tensor ${\widetilde{\check{\mathcal{T}}}_{M}{}}^{\mu}{}_{\nu}$ represent that it is raised and lowered with the metric $g_{\mu\nu}$. With these tensors we can construct the single and zeroth copies decoupled sources as:
\begin{align}
    \nonumber {J_{M}^{+}}^{\mu} \equiv -\frac{2}{\alpha}K^{\nu} {T_{M}^{+}}^{\mu}{}_{\nu} \, , \quad {J_{M}^{-}}^{\mu} \equiv -\frac{2}{\beta}\mathcal{K}^{\nu}{T_{M}^{-}}^{\mu}{}_{\nu} \, , \quad  {j_{M}^{+}} \equiv -\frac{2}{\alpha} K_{\mu}{J_{M}^{+}}^{\mu} \, , \quad {j_{M}^{-}} \equiv -\frac{2}{\beta}\mathcal{K}_{\mu}{J_{M}^{-}}^{\mu} \, .
\end{align}
The parameter $\widehat{m}$ in equations (\ref{decoup_eqns}) is defined in terms of the Fierz-Pauli mass, i.e., the flat-space graviton mass $m$, 
\begin{align} 
    \widehat{m}^{2}= \frac{2m^{2}P_{0}}{C\kappa^{2}}(C^{2} {\kappa_{g}}^{2}+{\kappa_{f}}^{2})\, . 
\end{align}
It is important to note that even if the flat-space graviton mass $m$ is nonzero, we can have zero mass modes by imposing the interaction constant $P_{0}$ to be zero. In the following subsections we study particular examples under this formalism.

\subsection{AdS waves effectively coupled to matter}

AdS waves are exact gravitational waves propagating on an AdS background and were studied in bigravity 
in \cite{Ayon-Beato:2018hxz}.  The line element for AdS waves in Poincar\'e coordinates can be written in Kerr-Schild form and is specified by
\begin{align}
    \nonumber d\overline{s}^2=\frac{l^2}{y^2}\left(-2dudv+dx^2+dy^2\right) \, , \quad k_{\mu}dx^{\mu}=-\frac{l}{y}du \, , 
\end{align}
where the scalar function that appears on the ansatz will depend on the wave solution. We will use the separable solutions presented by the authors in \cite{Ayon-Beato:2018hxz} where, depending on the nature of $\widehat{m}$, one can have massive $\widehat{m}\ne 0$ or massless profiles $\widehat{m} = 0 $.  

The separable solutions presented in \cite{Ayon-Beato:2018hxz} for the massless profiles where  $P_{0} = 0$ give rise to the following functions for the ansatz:
\begin{align}
    \nonumber &\phi^{h}_{g}(u,x,y) =-\left(\frac{\kappa_{g}}{2}\right) \frac{2}{{\kappa_{g}}^{2}}f_{3}(u)\left(\frac{y}{l}\right)^{3},  \\
    \nonumber &\mathsf{\phi}^{h}_{f}(u,x,y) =-\left(\frac{\kappa_{f}}{2}\right) \frac{2}{{\kappa_{f}}^{2}}\left[h_{3}(u)\left(\frac{y}{l}\right)^{3}+\frac{h_{2}(u)}{l^{2}}\left(x^{2}+y^{2}\right)+h_{1}(u)\frac{x}{l}+h_{0}(u) \right]\, ,
\end{align}
where  $h_{0}, h_{1},h_{2}, h_{3}$ and $f_{3}$ are arbitrary functions. The massive profiles $P_{0}\ne 0$ obtained in \cite{Ayon-Beato:2018hxz} for the ansatz are:
\begin{align}
    \label{ads_hom_massive} 
    \nonumber &\phi^{h}_{g}=-\left(\frac{\kappa_{g}}{2}\right)\frac{2}{{\kappa_{f}}^2+C^{2}{\kappa_{g}}^{2}}\left[\frac{\kappa^{2}}{{\kappa_{g}}^{2}} f_{3}(u)\left(\frac{y}{l}\right)^{3}\right.\left.-\,C^{2} \left(h_{1}(u)\left(\frac{y}{l}\right)^{\,\rho_{+}}+h_{2}(u)\left(\frac{y}{l}\right)^{\,\rho_{-}}\right)\right], \\
    &\phi^{h}_{f} =-\left(\frac{\kappa_{f}}{2}\right)\frac{2}{{\kappa_{f}}^{2}+C^{2}{\kappa_{g}}^{2}}\left[\frac{\kappa^{2}}{{\kappa_{f}}^{2}} f_{3}(u)\left(\frac{y}{l}\right)^{3}\right.\left.+\left(h_{1}(u)\left(\frac{y}{l}\right)^{\,\rho_{+}}+h_{2}(u)\left(\frac{y}{l}\right)^{\,\rho_{-}}\right)\right] \, ,
\end{align}
where $h_{1},h_{2}$ and $f_{3}$ are arbitrary functions and 
\begin{align}
    \nonumber \rho_{\pm}\equiv\frac{3}{2}\pm l \sqrt{\widehat{m}^{2}-{m_{BF}}^{2}}\, , \quad {m_{BF}}^{2}\equiv -\frac{9}{4l^{2}} \, . 
\end{align}
Here the roots $\rho_{\pm}$ determine the linearly independent
solutions of the ordinary Euler equation for the massive modes. ${m_{BF}}^{2}$ is the Breitenlohner-Freedman bound which gives the lowest possible value that a stable scalar field can take on an AdS background \cite{Breitenlohner:1982jf}. The massless and both massive vacuum cases presented here for AdS waves i.e., the saturated and non-saturated solutions, satisfy the equations in (\ref{time_dep_eom}).

Using the redefinitions in (\ref{redef_ads_2}) we can decouple the profiles and obtain the massive and massless modes,
\begin{align}
    \nonumber \phi^{h +}(u,x,y)= f_{3}(u)\left(\frac{y}{l}\right)^{3} \, ,  \quad \phi^{h -}(u,x,y) =h_{1}(u)\left(\frac{y}{l}\right)^{\,\rho_{+}}+h_{2}(u)\left(\frac{y}{l}\right)^{\,\rho_{-}} \, ,
\end{align}
where, the masses of the decoupled zeroth copy fields, i.e. the masses of the scalar profiles $\phi^{+}$ and $\phi^{-}$ are given by
\begin{align} 
    \nonumber  {m^{+}_{Z}}^{2} = 0\, , \quad   {m^{-}_{Z}}^{2} = \widehat{m}^{2}\, , 
\end{align}
The decoupled double, single and zeroth copy fields made for this vacuum AdS waves satisfy the equations in (\ref{decoup_eqns}) with the respective masses and the coupled fields satisfy the equations in (\ref{time_dep_eom}) as shown in \cite{Garcia-Compean:2024uie}. We will use these homogeneous solutions to construct the full solution which will be coupled to matter. In what follows we will consider first an scalar field as the matter content and later we will consider the Maxwell field.

\subsubsection{AdS waves in bigravity coupled to a scalar field}

In this example we consider the AdS solution coupled to a scalar field $\Phi$. Using the effective metric ${g_{E}}^{\mu\nu}$, the Lagrangian and the energy-momentum tensor for an scalar field are given by:
\begin{align}
    \nonumber
    \mathcal{L}_{M}=-\frac{1}{2}{g_{E}}^{\mu\nu}\partial_{\mu}\Phi\partial_{\mu}\Phi \, , \quad {T_{E}}_{\mu\nu}=\partial_{\mu}\Phi\partial_{\nu}\Phi-\frac{1}{2}{g_{E}}_{\mu\nu}\,{g_{E}}^{\rho\sigma}\partial_{\rho}\Phi\partial_{\sigma}\Phi \, .
\end{align}
The dynamics of the scalar field $\Phi$ is governed by the wave equation, where the solution of the scalar field is an arbitrary function of the retarded time
\begin{align}
    \nonumber
    {g_{E}}^{\mu\nu} \partial_{\mu}\partial_{\nu}\Phi = 0 \quad \Rightarrow \quad \Phi = \Phi (u)\, .
\end{align}
In order to obtain these inhomogeneous functions we use the decoupled fields obtained in \cite{Ayon-Beato:2018hxz} 
\begin{align}
    \label{ads_scalar}
    \nonumber {\phi^{+}}^{i}   = &\, -\frac{\kappa_{g}\kappa_{f}(\alpha+\beta C)^{2}}{2\kappa}\dot{\Phi}^{2}y^{2}\, , \\ 
    {\phi^{-}}^{i}  = & \, 
        \begin{cases}
            -\dfrac{(\beta{\kappa_f}^{2}-\alpha C{\kappa_g}^2)(\alpha+\beta C)}{C(\hat{m}^2 l^2+2)}\dot{\Phi}^{2} y^2\, \quad &\widehat{m}^2 \neq {m_{sr}}^2,\\
            \dfrac{(\beta{\kappa_f}^{2}-\alpha C{\kappa_g}^{2})(\alpha+\beta C)}{C}\dot{\Phi}^2y^2\ln\left(\frac{y}{l}\right),&\widehat{m}^2=m_\mathrm{sr}^2,
    \end{cases}
\end{align}
where the scalar source inhomogeneity enters in resonance with the vacuum modes when the mass $\hat{m}^{2}$ becomes $m_{sr}^{2}$,
\begin{align}
    \nonumber
    m_{sr}^{2}\equiv-\frac{2}{l^{2}} =m_{BF}^2+\frac{1}{4 l ^2} \, .
\end{align}
Even though this mass is negative, this case describes physically admissible configurations as the mass is above the Breitenlohner-Freedman bound. By using (\ref{ads_scalar}) in (\ref{redef_invs}) we can construct the scalar fields that appear in the metric, $\phi_{g}$ and $\phi_{f}$, which will be a sum of the homogenous and inhomogenous solutions. We find that both massive profiles in (\ref{ads_scalar}) satisfy the equations (\ref{time_dep_eom}) and the complete decoupled fields $\phi^{+}$ and $\phi^{-}$ satisfy the decoupled equations (\ref{decoup_eqns}).

\subsubsection{AdS waves in bigravity coupled to the electromagnetic field}
We can also consider the case of AdS waves with the two fields $g_{\mu\nu}$ and $f_{\mu\nu}$ coupled to the Maxwell field $A_{M\,\mu}$. The electromagnetic field Lagrangian and energy-momentum tensor are given by:
\begin{align}
    \nonumber \mathcal{L}_{M}=-\frac{1}{16\pi}{g_{E}{}}^{\mu\rho}{g_{E}{}}^{\nu\sigma}F_{\mu\nu}F_{\rho\sigma}\,, \quad {T_{E}{}}_{\mu\nu}=\frac{1}{4 \pi}\left({g_{E}}^{\rho\sigma}F_{\mu\rho}F_{\nu\sigma}-\frac{1}{4}{g_{E}}_{\mu\nu}{g_{E}{}}^{\gamma\rho}{g_{E}{}}^{\delta\sigma}F_{\gamma\delta}F_{\rho\sigma}\right) \,.
\end{align}
In \cite{Ayon-Beato:2018hxz} it is shown, after a gauge fixing procedure, that the most general Maxwell potential ${A_{M}}_{\mu}$ compatible with the pure radiation constraints, i.e., supporting the AdS waves, is proportional to the null rays $k_{\mu}$ and is given by
\begin{align}
A_{M\,\mu}=\left(A_{M}^{(u)},0,0,0\right) \, .
\end{align}
This choice of the potential reduces the Maxwell equations to the harmonic equation
\begin{align}
    \Delta\, A_{M}^{(u)} = 0,
\end{align}
where $\Delta \equiv \partial_{x}^{2}+\partial_{y}^{2}$. The general solution is the real part of a holomorphic function on the complex wavefront coordinate $z=x+iy$,
\begin{align}
    \nonumber A_{M}^{(u)}(u,x,y)=\partial_{z} a(u,z)+\overline{\partial_{z}a(u,z)}\, , 
\end{align}
which, as shown in \cite{Ayon-Beato:2018hxz}, can be written in terms of the derivatives of other holomorphic function $a(u,z)$. We will study the massless $\widehat{m}=0$ and massive $\widehat{m}\ne0$ sectors of the solution. 

For the inhomogeneous decoupled fields on the massless profiles sector, $\hat{m}=0$, we have:
\begin{align}
&\phi^{+\, i} =\frac{{\kappa_{g}{}}^{2}{\kappa_{f}{}}^{2}}{16\pi\kappa^{2} l^{2}}y^{3}\Delta \left(\frac{a\,\overline{a}}{y}\right),  \\
&\phi^{-\, i}=\frac{(\beta{\kappa_{f}}^{2}-\alpha C{\kappa_{g}}^{2})}{16\pi l^{2}C(\alpha+\beta C)}y^{3}\Delta\left(\frac{a\,\overline{a}}{y}\right),\quad\widehat{m}^{2}=0\, . 
\end{align}
With these fields, one can contract the bigravity equations with Killing vectors fields to obtain that the massless sector of the solution satisfy the equations (\ref{time_dep_eom}) and (\ref{decoup_eqns}).

Assuming the Maxwell field is compatible with the spatial translations $\partial_{x}$, is possible to consider particular solutions in the massive sector, where $\widehat{m}\ne0$. This symmetry implies that one can write:
\begin{align}
    a(u,z)=\frac{1}{2}D_{2}z^{2}+D_{1}z+D_{0},
\end{align}
where $D_{0}=D_{0}(u)$, $D_{1}=D_{1}(u)$ and $D_{2}=D_{2}(u)$. $D_{1}$ and $D_{0}$ can be eliminated by a residual transformation \cite{Ayon-Beato:2018hxz}. Then, the inhomogeneous fields that contribute to the decoupled AdS-wave profiles in the massive case $\widehat{m}\ne0$ are given by
\begin{align}
    \label{ads_em_massive}
    \nonumber &\phi^{+\, i} =\frac{{\kappa_{g}{}}^{2}{\kappa_{f}{}}^{2}}{2\pi\kappa^{2} l^{2}}|D_{2}|^{2}y^{4},  \\
    &\phi^{-\, i} =
        \begin{cases}
            -\dfrac{2(\beta{\kappa_{f}{}}^{2}-\alpha C{\kappa_{g}{}}^{2})}{\pi l^2 C (\alpha+\beta C)(l^2\widehat{m}^2-4)}y^4,& \widehat{m}^2\neq {m_{er}{}}^{2}\, , 
            \\\dfrac{2(\beta{\kappa_{f}{}}^{2}-\alpha C {\kappa_{g}{}}^{2})|D_2|^2}{5\pi l^2 C (\alpha+\beta C)}y^4\ln\frac{y}{l},& \widehat{m}^2={m_{er}{}}^2\, .
        \end{cases} 
\end{align}
The inhomogeneity enters in resonance with one of the vacuum modes (\ref{ads_hom_massive}) when the mass takes the value:
\begin{align}
    \nonumber m^{2}_{\text{er}}=\frac{4}{l^{2}}.
\end{align}
The massive sector of the solution also satisfies the equations (\ref{time_dep_eom}) and (\ref{decoup_eqns}). Then, with these results is possible to see that these time-dependent solutions are contained within the formalism presented in the previous sections.

\clearpage
\section{Pleba\'nski-Demia\'nski double Kerr-Schild solutions in bigravity and the double copy}\label{PD in Bigravity}

Pleba\'nski-Demia\'nski solutions are a general class of  solutions of Einstein-Maxwell equations and depend on seven parameters, which are mass, NUT parameter, two kinematical parameters, electric and magnetic charge and a cosmological constant. These solutions can be written in a double Kerr-Schild form after an analytical continuation of the parameters, and the known solutions such as Schwarzschild, Kerr and its generalizations with charge and cosmological constants are obtained at certain limits \cite{plebanski1975class,plebanski1976rotating}.

The Kerr-Schild single copy of some black hole solutions were obtained in \cite{Monteiro:2014cda,Carrillo-Gonzalez:2017iyj}, as well as in \cite{Luna:2015paa} for the Taub-NUT solution and the Kerr-Taub-NUT-(A)dS solution in \cite{Farnsworth:2023mff}, considering the Newman-Penrose map. In order to study the Pleba\'nski-Demia\'nski solution, we need to consider an external Maxwell field. As commented before, even though coupling matter from the gravity side has its complication at the level of scattering amplitudes, we can study the classical equations these fields satisfy. 

In \cite{Garcia-Compean:2024uie}, for the single copy of the KN-(A)dS solution in bigravity, there are some terms that arise when coupling matter to the solutions, which differ from the Maxwell solution coupled to the external source and were discussed in GR in \cite{Bah:2019sda,Alkac:2021bav}. In this work this complication is not presented, as we note that this additional contributions are zero in Pleba\'nski coordinates. A key characteristic of these coordinates is that they permit a double KS form and linearize the Ricci tensor with mixed indices. We will study a bigravity generalization of the Pleba\'nski-Demia\'nski solutions in the context of the double Kerr-Schild double copy and analyze the equations of motion at the double, single and zeroth copy level, as was done in as in \cite{Carrillo-Gonzalez:2017iyj} for GR. In order to do this, we first consider the double copy relations for the Pleba\'nski-Demia\'nski solutions in GR and then consider this solution in massive bigravity.

\subsection{Pleba\'nski-Demia\'nski solution in GR}

The goal of this section is to work with a double Kerr-Schild ansatz in bigravity, extending the work in  \cite{Garcia-Compean:2024uie}. First, we examine the double copy of the Taub-NUT solution \cite{Taub:1950ez, Newman:1963yy}, which was studied in \cite{Luna:2015paa} and whose single copy is
a dyon with the mass related to the electric charge and the NUT charge related to the magnetic charge of the gravity and gauge theory respectively. This solution was also studied in \cite{Luna:2018dpt} using the Weyl double copy. After analyzing the solution in GR, we will proceed to present this solution in bigravity.

\subsubsection{Taub-NUT Kerr-Schild double copy in GR}

The Taub-NUT solution permits a double Kerr-Schild form when written in Pleba\'nski coordinates \cite{plebanski1975class}. The line element of a metric written in a double Kerr-Schild form is:
\begin{align*}
    d s^2=d\overline{s}^2+\phi \left( k_{\mu} dx^{\mu}\right)^{2}+\psi \left(l_{\mu} dx^{\mu}\right)^{2}\, .
\end{align*}
The metric can be written as:
\begin{align*}
    \nonumber g_{\mu\nu}= \overline{g}_{\mu\nu}+\kappa h_{\mu\nu} \, ,  \quad h_{\mu\nu} = \phi k_{\mu}k_{\nu} +\psi l_{\mu}l_{\nu} \, , 
\end{align*}
As discussed in \cite{Luna:2015paa}, in Pleba\'nski coordinates $(\tilde{\tau},\tilde{\sigma},p,q)$ the background for the spacetime is given by
\begin{align}
    \label{Pleb_bg}
    \nonumber d\overline{s}^2&=-\frac{1}{q^2-p^2}\left[\overline{\Delta}_p(d\tilde{\tau}+q^2d\tilde{\sigma})^2+\overline{\Delta}_q(d\tilde{\tau}+p^2d\tilde{\sigma})^2\right]-2(d\tilde{\tau}+q^2d\tilde{\sigma})dp-2(d\tilde{\tau}+p^2d\sigma)dq \, , \\
    \overline{\Delta}_p&\equiv\gamma-\epsilon p^2+\frac{\Lambda}{3} p^4,\quad\overline{\Delta}_q\equiv-\gamma+\epsilon q^2-\frac{\Lambda}{3} q^4\, . 
\end{align}
with the following scalar functions and null vectors written as
\begin{align}
    \label{null vects_gr}
    \phi=\frac{\kappa}{2}\frac{2mq}{8 \,\pi \,(q^2-p^2)} \, , \quad l_\mu=(1,p^2,0,0) \, , \qquad \psi=\frac{\kappa}{2}\frac{2Np}{8 \, \pi \, (q^2-p^2)}\, ,\quad k_\mu=(1,q^2,0,0)\, ,
\end{align}
Generically, in the double Kerr-Schild ansatz, $R^{\mu}{}_{\nu}$ is not linear in the perturbation. However, in these specific coordinates, non-linear contributions to the Ricci tensor vanish. 

In this solution, $m$, $N$, $\epsilon$, $\gamma$ and $\Lambda$ are parameters of the solution. These free parameters can be interpreted as the mass, the NUT parameter, kinematical parameters and cosmological constant when considering certain limiting cases and transformations of the metric. Via the Kerr-Schild double copy procedure, the single and zeroth copy fields are:
\begin{align*}
    A_\mu^a=c^a\left(\phi k_\mu+\psi l_\mu\right) \,, \quad \Upsilon^{aa'}=c^a\tilde{c}^{a'} (\phi+\psi) \,,
\end{align*} 
whose colored-stripped Abelian fields satisfy their corresponding equations:
\begin{align*}
     \overline{\nabla}_{\lambda}F^{\lambda\mu} = \, 0 \, , \qquad \left(\overline{\nabla}^{2}-\frac{\overline{R}}{6}\right)\Upsilon = \, 0 \,, 
\end{align*}
where $F_{\mu\nu}\equiv \,  \overline{\nabla}_{\mu} A_{\nu}-\overline{\nabla}_{\nu}A_{\mu}$ and $\Upsilon \equiv \phi + \psi$. In this vacuum case, we only have the homogeneous scalar function, $\Upsilon=\Upsilon^{h}$. We use the superscript ``h'' for the homogeneous (vacuum) solution, which will consider the mass and NUT parameters, and will be using ``i'' for the inhomogeneous solution which for the case when is coupled to matter, so that we consider the electric and magnetic charge parameters, whereas the full solution is the sum of both terms, $\Upsilon=\Upsilon^{h}+\Upsilon^{i}$. As interpreted in \cite{Luna:2015paa}, the single copy of the Taub-NUT solution is a dyon, which has electric  and magnetic charge related to the mass and NUT parameters ($m \rightarrow q_{gauge}, N \rightarrow g_{gauge}$). The dyon interpretation coincides with the Weyl double copy result \cite{Luna:2018dpt}, which fixes a prefactor on the scalar function.

\subsubsection{Pleba\'nski-Demia\'nski Kerr-Schild double copy in GR}
Our purpose in this work is not only to consider these solutions, i.e., Taub-NUT in bigravity, but also to work with a generalization of this solution, the Pleba\'nski-Demia\'nski solution, which is coupled to matter, so that it considers electric $Q$ and magnetic $G$ charges \cite{plebanski1976rotating}. Using the background element in (\ref{Pleb_bg}) and null vectors in (\ref{null vects_gr}), the scalar functions $\phi$ and $\psi$ take the form
\begin{align}
    \label{Pleb_null_gr}
    \phi=\frac{\kappa}{2}\frac{2mq-{Q}^{2}}{8 \, \pi\, (q^2-p^2)} \, , \quad l_\mu=(1,p^2,0,0) \, , \qquad \psi=\frac{\kappa}{2}\frac{2Np+{G}^2}{8\,\pi(q^2-p^2)}\, ,\quad k_\mu=(1,q^2,0,0)\, ,
\end{align}
and the electromagnetic field is:
\begin{align*}
    A_{\text{M}\,\mu}=\frac{1}{q^{2}-p^{2}}\big(Q \,q \,l_{\mu}+ G\, p \,k_{\mu}\big) \, , 
\end{align*}
with the usual definitions for the electromagnetic energy-momentum and field strength tensors,
\begin{align}
    \nonumber F_{M\,\mu\nu}\equiv &\, {\nabla}_{\mu} A_{M\, \nu}-{\nabla}_{\nu}A_{M\, \mu}\,= \overline{\nabla}_{\mu} A_{M\, \nu}-\overline{\nabla}_{\nu}A_{M\, \mu}\, , \qquad{\nabla}_{\mu}F_{M}^{\mu\nu}=0\, , \\
    \nonumber {T_{M}}^{\, \mu}{}_{\nu}\equiv &\,  g^{\mu\alpha}{T_{M}}_{\,\alpha\nu}=\frac{1}{4\,\pi} \left({F_{M}}^{\,\mu\alpha}F_{M\, \nu\alpha}-\frac{1}{4}F_{M\,\alpha\beta}F_{M}{}^{\alpha\beta}\,\delta^{\mu}{}_{\nu} \right)\, , \quad T_{M}\equiv {T_{M}}^{\, \mu}{}_{\mu}=0\, ,
\end{align}
which specifies the desired solution. The parameters $Q$ and $G$ can be associated to the electric and magnetic charge in specific transformations of the solution. The Taub-NUT solution studied in \cite{Luna:2015paa} is a vacuum solution and is recovered in the non-rotating case, $\gamma=0, \epsilon=1$ and by setting $Q=0=G$. Pleba\'nski-Demia\'nski solutions in GR were also studied in the context of the Weyl double copy \cite{Easson:2022zoh}. Here we will apply the Kerr-Schild double copy procedure as in \cite{Carrillo-Gonzalez:2017iyj}, so that by contracting with timelike Killing vectors, the equations of motion for this Pleba\'nski-Demia\'nski solution in terms of the double, single and zeroth copy fields are:
\begin{align}
    \label{g_evac_eom}
    \nonumber \overline{\mathcal{E}}(h^{\mu}{}_{\nu}) - \frac{\overline{R}}{4} \, h^{\mu}{}_{\nu} \,  = \,  \frac{\kappa}{2} {\check{T}_{M}{}}^{\mu}{}_{\nu}\, & , \quad  h_{\mu\nu}=\phi k_{\mu}k_{\nu}+\psi l_{\mu}l_{\nu} \, ,\\
    \nonumber \overline{\nabla}_{\lambda}F^{\lambda\mu}\, = \, g {J_{M}}^{\mu}\, & , \quad A_{\mu}= \phi k_{\mu}+\psi l_{\mu},\\
    \overline{\nabla}^{2}\Upsilon -\frac{\overline{R}}{6}(\Upsilon+\Upsilon^{i}) \,  =  \,  y\,j_{M} \, &, \quad \Upsilon=\phi+\psi\, , 
\end{align}
where we have source terms and $\Upsilon^{i}$ is the sum of the electric and magnetic terms of the scalar functions. These are the equations of motion for the double copy fields defined for this solution, and by considering vanishing charges, we recover \cite{Luna:2015paa}. The double Kerr-Schild nature of the solution is in the definition of each of the fields as the sum of two terms. In the case $G=Q$, we obtain ${T_{M}}^{\, \mu}{}_{\nu}=0$.

Remarkably, in this coordinates, the single and zeroth copy source terms in (\ref{g_evac_eom}) for the Pleba\'nski-Demia\'nski solution can be written as:
\begin{align}
    \label{sources_gr}
    \nonumber g\,{J_{M}}^{\mu} & =  - \frac{4}{\left(q^{2}-p^{2}\right)^{3}} \Upsilon^{i}\left( \frac{1}{2}\left(q^{2}+p^{2}\right),-1,0,0 \right)\, , \\
        y\,j_{M} & = -\frac{2\big(\overline{\Delta}_{p}-\overline{\Delta}_{q}\big)}{\left(q^{2}-p^{2}\right)^{2}} \Upsilon^{i} \, , \quad \Upsilon^{i}=\frac{-{Q}^{2}+G^{2}}{q^2-p^2}
\end{align}
Then, for the zeroth copy, we note that the inhomogeneous and homogeneous functions satisfy the equations:
\begin{align}
\label{inhom_eqs_gr}
    \left(\overline{\nabla}^2-\frac{\overline{R}}{6}\right)\Upsilon^{h}=0 \, , \qquad
    \left(\overline{\nabla}^2-\frac{\overline{R}}{3}+\frac{2\big(\overline{\Delta}_{p}-\overline{\Delta}_{q}\big)}{\left(q^{2}-p^{2}\right)^{2}}\right)\Upsilon^{i} =0 \, .
\end{align}
The homogeneous solution satisfies the same equation as before, and the inhomogeneous solution has an additional term which depends on the coordinates. The analog equation for the Reissner-N\"ordstrom solution in the context of bigravity is presented in \cite{Garcia-Compean:2024uie}, where we reduced to the non-rotating case in global static coordinates. Equation (\ref{inhom_eqs_gr}) considers the scalar functions that appear in the the metric for the rotating case with non-vanishing magnetic and NUT parameter in Pleba\'nski coordinates. The fields $\Upsilon^{i}$ obey a Klein-Gordon equation with a scalar curvature and a term with a coefficient which depends on the spatial coordinates $p$ and $q$ and also on the kinematical parameters and cosmological constant. The simplicity of this equations may be associated to the fact that, in this coordinates, the background metric has the kinematical information of the solution, and the deviation to the background depends on the dynamical parameters. 

We can recover some known solutions in spheroidal coordinates. For example, in the vacuum case, we have a relation of the coordinates $(\tilde{\tau},\tilde{\sigma},p,q)$ to the coordinates $(\tau,\sigma,p,q)$:
\begin{align}
    \label{transf_pleb}
    d\tilde{\tau}=d\tau+\frac{p^2dp}{\Delta_p}-\frac{q^2dq}{\Delta_q},\quad d\tilde{\sigma}=d\sigma-\frac{dp}{\Delta_p}+\frac{dq}{\Delta_q}\, , 
\end{align}
where 
\begin{align*}
    \Delta_p=\gamma-\epsilon p^2+\frac{\Lambda}{3} p^4-2Np,\quad\Delta_q=-\gamma+\epsilon q^2-\frac{\Lambda}{3} q^4-2mq \, , 
\end{align*}
and when $\Lambda=0$ and $N=0$, the coordinates $(\tau,\sigma,p,q)$ are related to the spheroidal coordinates $(t,r,\theta,\varphi)$ by:
\begin{align}
    \tau = t - a\varphi, \quad \sigma=-\frac{\varphi}{a},\quad q=r,\quad p=a\cos\theta \,,
\end{align}
which permit us recover the Kerr solution. We need to do an analytical continuation of the metric in the coordinates $(\tau,\sigma,p,q)$, which in the general case it means that we make $p \rightarrow ip, N \rightarrow - N  $ and $\gamma \rightarrow - \gamma$ and then apply (\ref{transf_pleb}) in order to arrive to the coordinates $(\tilde{\tau},\tilde{\sigma},p,q)$. Then, by considering certain limiting cases of the Pleba\'nski-Demia\'nski solution, the known type D black hole solutions e.g., the Schwarzschild and Kerr solutions, are recovered when returning to the real domain, as explained in \cite{plebanski1976rotating,Chong:2004hw}. Pleba\'nski coordinates are also useful from a computational standpoint, as the computation time in Maple was much shorter using these coordinates compared to spheroidal coordinates.

One task we will develop here is to generalize the work for KS double copy in bigravity in \cite{Garcia-Compean:2024uie} by using solutions coupled to matter. For this, we will use the Pleba\'nski-Demia\'nski solution which permits to consider a double Kerr-Schild ansatz formalism in bigravity in the context of the classical double copy relations.

\subsection{Stationary solution in bigravity}
We present a stationary double Kerr-Schild solution in bigravity written in Pleba\'nski coordinates. This solution satisfies relations analog to the ones in (\ref{waves_constraints}) but for stationary solutions: 
\begin{align}
\label{gf_evac_stat_constraints}
    \nonumber \frac{\overline{R}}{6}A^{\mu} + \frac{1}{\alpha}K^{\nu}W^{\mu}{}_{\nu} = 0 \, ,& \qquad \frac{1}{\alpha^2}K_{\mu}Z^{\mu} + \frac{\overline{R}}{6}\Upsilon_{g}\, = \, - \frac{\overline{R}}{6}\Upsilon^{i}_{g}\,, \\ \frac{\overline{\mathcal{R}}}{6}\mathcal{A}^{\mu} + \frac{1}{\beta}\mathcal{K}^{\nu}\mathcal{W}^{\mu}{}_{\nu} = 0 \,,& \qquad \frac{1}{\beta^2}\mathcal{K}_{\mu}\mathcal{Z}^{\mu} + \frac{\overline{\mathcal{R}}}{6}\Upsilon_{f}\, = \, - \frac{\overline{\mathcal{R}}}{6}\Upsilon^{i}_{f}\, .
\end{align}
We are modifying these relations with respect to the vacuum case by adding a term in the equation which has the fields $\Upsilon^{i}$, that depend on the electric and magnetic charges, in contrast to (\ref{waves_constraints}), where the vacuum and non-vacuum relations are the same. The results in (\ref{gf_evac_stat_constraints}) are written for Pleba\'nski coordinates, and will change if we transform to another frame of reference (e.g., Kerr-Newmann-(A)dS in global static coordinates as in \cite{Ayon-Beato:2015qtt, Garcia-Compean:2024uie}). 

Then, we can writedown the equations of motion for this stationary solution as
\begin{align}
    \label{gf_evac_eom}
    \nonumber \,^{(g)}\overline{\mathcal{E}}(h^{\mu}{}_{\nu}) - \frac{\overline{R}}{4} \, h^{\mu}{}_{\nu} \,  = \,  \frac{\kappa_{g}}{2} {\check{T}_{M}{}}^{\mu}{}_{\nu}\, & , \qquad \,^{(f)}\overline{\mathcal{E}}(\mathscr{h}^{\mu}{}_{\nu}) -\,\frac{\overline{\mathcal{R}}}{4} \, \mathscr{h}^{\mu}{}_{\nu} \,\, = \, \frac{\kappa_{f}}{2 \,C^{2}}{\check{\mathcal{T}}_{M}{}}^{\mu}{}_{\nu}\, ,\\
    \nonumber \overline{\nabla}_{\lambda}F^{\lambda\mu}\, = \, g_{g} J_{M}\, & , \qquad \overline{\nabla}_{\lambda}\mathcal{F}^{\lambda\mu} \, = \, g_{f} \mathcal{J}_{M} \, ,\\
    \overline{\nabla}^{2}\Upsilon_{g} -\frac{\overline{R}}{6}(\Upsilon_{g}+\Upsilon^{i}_{g}) \,  =  \,  y_{g}\,j_{M} \, &, \qquad \,^{(f)}\overline{\nabla}^{2}\Upsilon_{f} -\frac{\overline{\mathcal{R}}}{6}(\Upsilon_{f}+\Upsilon^{i}_{f}) \, = \, y_{f}\,\mathcal{j}_{M} \, .
\end{align}
These results are consistent with what we obtained in \cite{Garcia-Compean:2024uie}, with the consideration that we worked in global static coordinates as used in \cite{Ayon-Beato:2018hxz,Carrillo-Gonzalez:2017iyj}. In global static coordinates, the sources $J_{A\,\mu}, \mathcal{J}_{A\,\mu}$ defined in \cite{Garcia-Compean:2024uie} do not vanish for the rotating case, and we have an extra contribution to the equations of motion which cannot be expressed in terms of our fields (\ref{full_fields}), which was also presented in \cite{Alkac:2021bav}. In Pleba\'nski coordinates we do not encounter the problem of this term, and we are left with equations of motion whose interpretation in terms of the fields is more transparent. The results in (\ref{gf_evac_eom}) for the spin-2 interacting fields are also consistent with the multi-metric gravity results \cite{Wood:2024acv} for the case of two metrics and separate matter sectors.

In (\ref{g_evac_eom}) we provided information about the double copy of the Pleba\'nski-Demia\'nski solutions in GR and now we will proceed to present the solution for bigravity. We consider the decoupled matter sector and the sector coupled via the effective metric, as well as some specific cases of these solutions.

\subsection{Pleba\'nski-Demia\'nski solution in bigravity}
The Pleba\'nski-Demia\'nski metric in GR \cite{plebanski1975class,plebanski1976rotating} can be written in double Kerr-Schild form. We can consider a similar family in solutions in bigravity using the ansatz (\ref{ansatz_double ks bg2}). To specify the solution in bigravity, we will use the same maximally symmetric background and the same null vectors as used in GR, (\ref{Pleb_bg}) and (\ref{Pleb_null_gr}), 
\begin{align}
    \label{Pleb_bigrav_bg}
    \nonumber d\overline{s}^2&=-\frac{1}{q^2-p^2}\left[\overline{\Delta}_p(d\tilde{\tau}+q^2d\tilde{\sigma})^2+\overline{\Delta}_q(d\tilde{\tau}+p^2d\tilde{\sigma})^2\right]-2(d\tilde{\tau}+q^2d\tilde{\sigma})dp-2(d\tilde{\tau}+p^2d\sigma)dq \, , \\
    \nonumber\overline{\Delta}_p&\equiv\gamma-\epsilon p^2+\lambda p^4,\quad\overline{\Delta}_q\equiv-\gamma+\epsilon q^2-\lambda q^4\, ,\\
    l_\mu&=\mathscr{l}_{\mu}=(1,p^2,0,0) \, , \quad k_\mu=\mathscr{k}_\mu=(1,q^2,0,0)\, ,
\end{align}
The  scalar functions for each metric have the same functional form as in (\ref{Pleb_null_gr}):
\begin{align}
    \label{scalar_matter_indep}
    \nonumber \phi_{g}= \frac{\kappa_{g}}{2}\frac{2m_{1}p-{Q_{1}}^2}{8\,\pi\,(q^2-p^2)}\, , \quad \psi_{g}=\frac{\kappa_{g}}{2}\frac{2N_{1}q+{G_{1}}^{2}}{8\,\pi\,(q^2-p^2)}\, , \\
    \phi_{f}=\frac{\kappa_{f}}{2}\frac{2m_{2}p-{Q_{2}}^2}{8\,\pi\,(q^2-p^2)}\, , \quad \psi_{f}=\frac{\kappa_{f}}{2}\frac{2N_{2}q+{G_{2}}^{2}}{8\,\pi\,(q^2-p^2)}\, , 
\end{align}
where $m_{1}$ is the mass parameter of the metric $g_{\mu\nu}$ and $m_{2}$ for $f_{\mu\nu}$ and so on. In order to define the matter content of the solution, we will consider the case of independent matter coupling and matter coupling via the effective metric, so that we have these two options for Pleba\'nski-Demia\'nski solutions in bigravity. If couple matter independently, we can have two distinct Maxwell fields for each sector, $A_{M\,\mu}, \mathcal{A}_{M\,\mu}$, and when using the effective metric we will have only one Maxwell effective field, $A^{E}_{M\,\mu}$. We will study each case briefly.

\subsection{Separate matter sectors}
Let us consider the case where we couple the two metrics independently to two electromagnetic fields. Then, we will have electrovacuum for both metrics $g_{\mu\nu}$ and $f_{\mu\nu}$. We can consider that the matter sectors are decoupled, i.e., 
\begin{align}
    \nonumber S_{\text{Mat}}[g,f,A_{M},\mathcal{A}_{M}]\equiv \, & S_{\text{Mat}}[g,A_{M}] +  S_{\text{Mat}}[f, \mathcal{A}_{M}]\, ,
\end{align}
so that we have two electromagnetic fields $A_{M\,\mu}$ and $\mathcal{A}_{M\,\mu}$. The field $A_{M}$ couples only to $g_{\mu\nu}$ and $\mathcal{A}_{M}$ only to $f_{\mu\nu}$. In this sense, we are coupling separate matter sectors to the metrics.

For this solution, the two Maxwell fields take the form:
\begin{align}
    \label{matter_indep}
    A_{\text{M}\,\mu}=\frac{1}{q^{2}-p^{2}}\big(Q_{1} \,q \,l_{\mu}+ G_{1}\, p \,k_{\mu}\big) \, ,\quad \mathcal{A}_{\text{M}\,\mu}=\frac{C}{q^{2}-p^{2}}\big(Q_{2} \,q \,l_{\mu}+ G_{2}\, p \,k_{\mu}\big) \, ,
\end{align}
With these fields we construct the expression for the field strength tensors $F_{M\,\mu\nu}, \mathcal{F}_{M\,\mu\nu}$ and the energy-momentum tensors $T_{M\,\mu\nu}, \mathcal{T}_{M\,\mu\nu}$ as follows
\begin{align}
    \nonumber F_{M\,\mu\nu}\equiv &\,  \,^{(g)}\overline{\nabla}_{\mu} A_{M\, \nu}-\,^{(g)}\overline{\nabla}_{\nu}A_{M\, \mu}\, , \qquad \,^{(g)}{\nabla}_{\mu}F_{M}^{\mu\nu}=0\, , \\
    \nonumber {T_{M}}^{\, \mu}{}_{\nu}\equiv &\,  g^{\mu\alpha}{T_{M}}_{\,\alpha\nu}=\frac{1}{4\,\pi} \left({F_{M}}^{\,\mu\alpha}F_{M\, \nu\alpha}-\frac{1}{4}F_{M\,\alpha\beta}F_{M}{}^{\alpha\beta}\,\delta^{\mu}{}_{\nu} \right)\, , \quad T_{M}\equiv {T_{M}}^{\, \mu}{}_{\mu}=0\, ,  \\ 
    \nonumber \mathcal{F}_{M\,\mu\nu}\equiv & \,  \,^{(f)}\overline{\nabla}_{\mu} \mathcal{A}_{M\, \nu}-\,^{(f)}\overline{\nabla}_{\nu}\mathcal{A}_{M\, \mu} \, , \qquad \,^{(f)}{\nabla}_{\mu}\mathcal{F}_{M}^{\mu\nu}=0 \, , \\
    \nonumber {\mathcal{T}_{M}}^{\, \mu}{}_{\nu}\equiv &\,  g^{\mu\alpha}{\mathcal{T}_{M}}_{\,\alpha\nu}=\frac{1}{4\,\pi} \left({\mathcal{F}_{M}}^{\,\mu\alpha}\mathcal{F}_{M\, \nu\alpha}-\frac{1}{4}\mathcal{F}_{M\,\alpha\beta}\mathcal{F}_{M}{}^{\alpha\beta}\,\delta^{\mu}{}_{\nu} \right)\, , \quad \mathcal{T}_{M}\equiv {\mathcal{T}_{M}}^{\, \mu}{}_{\mu}=0\, ,
\end{align}
where we have also included the dynamical equation for the matter strength fields. 

The solutions for bigravity given by (\ref{Pleb_bigrav_bg}), (\ref{scalar_matter_indep}) and (\ref{matter_indep}) satisfy (\ref{gf_evac_eom}) and depend on the mass parameters $m_{i}, N_{i}$, with $i=\{1,2\}$, where $m_{i}$ is the mass and $N_{i}$ the NUT parameters in each of the metrics, $\gamma$, $\epsilon$ are the  kinematical parameters, the electric and magnetic charge parameters $Q_{i}$, $G_{i}$ respectively and the cosmological constant $\Lambda$. For this Pleba\'nski-Demia\'nski solution with decoupled Maxwell sectors we can have different mass and charge parameters, but we have the same rotation parameter for both metrics and proportional cosmological constants.

We have that the sources for the single and zeroth copy in this solution are written as
\begin{align}
    \label{sources_bg}
    \nonumber g_{g}\,{J_{M}}^{\mu} & =  - \frac{4}{\left(q^{2}-p^{2}\right)^{3}} \Upsilon_{g}^{i}\left( \frac{1}{2}\left(q^{2}+p^{2}\right),-1,0,0 \right)\, , \\
    \nonumber g_{f}\,{\mathcal{J}_{M}}^{\mu} & =  - \frac{4}{C^{4}\left(q^{2}-p^{2}\right)^{3}} \Upsilon_{f}^{i}\left( \frac{1}{2}\left(q^{2}+p^{2}\right),-1,0,0 \right)\, , \\
    y_{g}\,j_{M} & = -\frac{2\big(\overline{\Delta}_{p}-\overline{\Delta}_{q}\big)}{\left(q^{2}-p^{2}\right)^{2}} \Upsilon_{g}^{i} \, , \qquad y_{f}\,\mathscr{j}_{M} = -\frac{2\big(\overline{\Delta}_{p}-\overline{\Delta}_{q}\big)}{C^{2}\left(q^{2}-p^{2}\right)^{2}} \Upsilon_{f}^{i} \, , 
\end{align}
so that, for the inhomogeneous scalar functions, the following hold:
\begin{align}
\label{inhom_eqs}
    \nonumber \overline{\nabla}^2\Upsilon^{i}_{g}-\frac{\overline{R}}{3}\Upsilon^{i}_{g}+\frac{2\big(\overline{\Delta}_{p}-\overline{\Delta}_{q}\big)}{\left(q^{2}-p^{2}\right)^{2}}\Upsilon^{i}_{g} =0 \, , \\ \,^{(f)}\overline{\nabla}^2\Upsilon^{i}_{f}-\frac{\overline{\mathcal{R}}}{3}\Upsilon^{i}_{f}+\frac{2\big(\overline{\Delta}_{p}-\overline{\Delta}_{q}\big)}{C^{2}\left(q^{2}-p^{2}\right)^{2}}\Upsilon^{i}_{f} =0 \, ,
\end{align}
which is analog to (\ref{inhom_eqs_gr}) in GR. The inhomogeneous decoupled scalar fields (\ref{redef_ads_2}) obey similar relations.

We can reduce to the case where we have an electrovacuum for the metric $g_{\mu\nu}$ and a vacuum for the metric $f_{\mu\nu}$ by setting the electric and magnetic charges of the metric $f_{\mu\nu}$ to zero, i.e., 
\begin{align*}
    Q_{2}=0=G_{2} \quad \Rightarrow \quad {\mathcal{T}_{M}}^{\mu}{}_{\nu}=0 \, . 
\end{align*}
The equations of motion for this case will be (\ref{gf_evac_eom}) with ${\mathcal{T}_{M}}^{\mu}{}_{\nu}=0$. The scalar functions which correspond to this solution are written as in (\ref{scalar_matter_indep}) with $ Q_{2}=0=G_{2}$ and we will have relations as in (\ref{gf_evac_stat_constraints}) with $\Upsilon^{i}_{f}=0$. If the charges of the metric $g_{\mu\nu}$ also vanish, $Q_{1}=0=G_{1}$, we have the vacuum stationary solutions, which in Pleba\'nski coordinates they satisfy (\ref{gf_evac_stat_constraints}) with $\Upsilon^{i}_{g}=0=\Upsilon^{i}_{f}$ and (\ref{gf_evac_eom}) with all the sources vanishing, i.e., ${T_{M}}^{\mu}{}_{\nu}=0={\mathcal{T}_{M}}^{\mu}{}_{\nu}$ and so on.

We can recover other massive gravity and GR solutions from the one presented here. We recover the bigravity solutions in \cite{Ayon-Beato:2015qtt} by setting $G_{i}=N_{i}$, $i={1,2}$, i.e., $\psi_{f}=0=\psi_{g}$ and by transforming to global static coordinates through the transformation in (\ref{transf_pleb}). We can reduce to the GR case in the weak gravity limit \cite{Babichev:2013pfa,Babichev:2010jd,Baccetti:2012bk} and obtain Pleba\'nski-Demia\'nski solutions \cite{plebanski1975class,plebanski1976rotating}. By setting $Q_{1}=0=G_{1}$ we recover the Kerr-NUT-(A)dS solution and in the non-rotating case reduces to the Taub-NUT solution in GR and the relations for the double copy equations (\ref{g_evac_eom}) without sources, as obtained in \cite{Luna:2015paa}.

\subsubsection{Proportional charges}
Now we study the case where the charges are equal or proportional. For example, let us consider the case where they are related as
\begin{align*}
    Q_{2}=\left(\frac{Q_{1}}{C}\right)  \, , \quad G_{2}=\left(\frac{G_{1}}{C}\right)  \, ,
\end{align*}
which implies that:
\begin{align}
    A_{\text{M}\,\mu}=\frac{1}{q^{2}-p^{2}}\big(Q_{1} \,q \,l_{\mu}+ G_{1}\, p \,k_{\mu}\big) = \mathcal{A}_{\text{M}\,\mu}.
\end{align}
Given the background metric (\ref{Pleb_bg}), null vectors (\ref{Pleb_null_gr}) and scalar functions (\ref{scalar_matter_indep}), this solution satisfies the equations in (\ref{gf_evac_eom}). Equivalently, we can consider $ Q_{2}={Q_{1}}, G_{2}={G_{1}}$ so that $A_{\text{M}\,\mu} = C \mathcal{A}_{\text{M}\,\mu}$. Then, by considering proportional or equal charges for solution in each metric, is equivalent to considering same or proportional fields $A_{\text{M}\,\mu}, \mathcal{A}_{\text{M}\,\mu}$ respectively.

\subsubsection{Proportional metrics}
We can consider a more restrictive case where $f_{\mu\nu}=C^{2}g_{\mu\nu}$. In order to do this, we can require the parameters to satisfy the relations:
\begin{align}
    \label{constraint_prop}
    m_{2}=\left(\frac{{\kappa_{g}}^{2}}{{\kappa_{f}}^{2}}\right) m_{1} \, , \quad N_{2}=\left(\frac{{\kappa_{g}}^{2}}{{\kappa_{f}}^{2}}\right) N_{1} \, , \quad Q_{2}=\left(\frac{{\kappa_{g}}}{{\kappa_{f}}}\right) Q_{1} \, , \quad G_{2}=\left(\frac{{\kappa_{g}}}{{\kappa_{f}}}\right) G_{1}  
\end{align}
which imply that 
\begin{align}
    \nonumber \mathscr{h}_{\mu\nu}= \left(\frac{\kappa_{g}}{\kappa_{f}}\right) h_{\mu\nu} \, , \quad \mathcal{A}_{\mu}= \left(\frac{\kappa_{g}}{\kappa_{f}}\right) A_{\mu} \, , \quad \Upsilon_{f}= \left(\frac{\kappa_{g}}{\kappa_{f}}\right) \Upsilon_{g} \, , \qquad \mathcal{A}_{\text{M}\,\mu} =\frac{C \kappa_{g}}{\kappa_{f}} A_{\text{M}\,\mu}\, .
\end{align}
The restriction for the parameters (\ref{constraint_prop}) along with the background metric (\ref{Pleb_bigrav_bg}), null vectors (\ref{Pleb_null_gr}), scalar functions and fields (\ref{scalar_matter_indep}), satisfies the equations of motion for bigravity in (\ref{gf_evac_eom}) with the caveat that the obtained equations at each level for both metrics are equivalent due to the proportionality relations the fields obey.

In this case of proportional metrics, if we decouple our equations (\ref{redef_ads_2}) to obtain the fields in the mass basis, we end up with the massive ``$-$'' fields to be vanishing, and the only field that propagates is the massless ``+'' field,
\begin{align}
    \Upsilon^{+}=-\frac{{\kappa_{g}{}}^{2}}{2\,} \frac{(C^{2}{\kappa_{g}}^{2}+{\kappa_{f}}^{2})}{8\,\pi\, \kappa^{2} \left(q^{2}-p^{2}\right)}\left(2m_{1}q-Q^{2}+2 N_{1} p+G^{2}\right) \, , \quad \Upsilon^{-} = 0 \, .
\end{align}

\subsection{Effective metric}

In this case, we only have one matter field, $A_{M\,\mu}\equiv A^{E}_{M\,\mu}$, and the energy-momentum tensors for each metric are related as in (\ref{eff em tensors}). In the decoupled matter sectors, we had electric $Q_{i}$ and magnetic $G_{i}$ parameters for both metrics. In this case, we found three solutions for the equations of bigravity, which restrict the parameter space for the electromagnetic charges. The effective coupled Maxwell field is given by
\begin{align*}
    A_{\text{M}\,\mu}=\frac{1}{q^{2}-p^{2}}\big(Q \,q \,l_{\mu}+ G\, p \,k_{\mu}\big)\, ,
\end{align*}
with electromagnetic parameters $Q$ and $G$ related to the electric and magnetic parameters of the metrics. We consider three cases for the effective metric formalism.

\subsubsection{The case $Q_{2}=0=G_{2}$}
We can express the case of electrovacuum in $g_{\mu\nu}$ and vacuum in $f_{\mu\nu}$ in the formalism of the effective metric by setting $\beta=0$, so that ${\mathcal{T}_{M}}^{\mu}{}_{\nu}=0$, i.e., we have 
\begin{align*}
    \beta=0\, , \quad g_{\text{E}\,\mu\nu}= \alpha^{2}g_{\mu\nu} \quad \Rightarrow \quad  {{T}_{E}}^{\mu}{}_{\nu} =\alpha^{2}\,{{T}_{M}}^{\mu}{}_{\nu} \ne 0 \, , \quad {\mathcal{T}_{M}}^{\mu}{}_{\nu} =0 \, .
\end{align*}
In terms of the effective metric, the energy-momentum tensor $T_{M\,\mu\nu}$ and the field strength tensor $F_{M\,\mu\nu}$ for the electromagnetic field can be written as
\begin{align}
    \label{em_eff_metric}
    \nonumber F_{M\,\mu\nu} & \equiv \,  \,^{(E)}\overline{\nabla}_{\mu} A_{M\, \nu}-\,^{(E)}\overline{\nabla}_{\nu}A_{M\, \mu}\, , \qquad \,^{(E)}{\nabla}_{\mu}F_{M}^{\mu\nu}=0\, , \\
    {T_{E}{}}_{\mu\nu}& \equiv \frac{1}{4 \pi}\left({g_{E}}^{\rho\sigma}F_{\mu\rho}F_{\nu\sigma}-\frac{1}{4}{g_{E}}_{\mu\nu}{g_{E}{}}^{\gamma\rho}{g_{E}{}}^{\delta\sigma}F_{\gamma\delta}F_{\rho\sigma}\right)\, , \quad {T_{E}}^{\, \mu}{}_{\nu} =\,  g_{E}^{\mu\alpha}{T_{E}}_{\,\alpha\nu}\, , \quad T_{E}=0\, ,  
\end{align}
where we have used the effective metric to raise and lower indices and $\,^{(E)}{\nabla}_{\lambda}$ is the covariant derivative associated to the effective metric, i.e., which is compatible with it, i.e. $\,^{(E)}{\nabla}_{\lambda} g_{E\,\mu\nu}=0$.  The equations of motion for this case are (\ref{gf_evac_eom}) with ${\mathcal{T}_{M}}^{\mu}{}_{\nu}=0$ and can be written using either ${{T}_{E}}^{\mu}{}_{\nu} $ or $ {{T}_{M}}^{\mu}{}_{\nu}$. 

\subsubsection{The case $G=Q$}
Let's considering a solution coupled via the effective metric, where the charges of both metrics are proportional. We can set the electric $Q$ and magnetic $G$ charge of the solution as equal, $G=Q$, if we consider the scalar functions in (\ref{scalar_matter_indep}) with the restrictions
\begin{align}
    \nonumber & Q_{1}=\sqrt{\frac{\alpha}{(C\beta+\alpha)}} Q\, ,  \quad Q_{2}= \sqrt{\frac{\beta}{C\left(C\beta+\alpha\right)}} Q\, , \\
    \nonumber &G_{1}=\sqrt{\frac{\alpha}{(C\beta+\alpha)}} G\, , \quad  G_{2} = \sqrt{\frac{\beta}{C\left(C\beta+\alpha\right)}} G \, , \quad G=Q\, . 
\end{align}
The fact that we restrict $G=Q$ lead us to
\begin{align}
    \label{equal_charges}
    A_{\text{M}\,\mu} = \frac{Q}{q^{2}-p^{2}}\big(q \,l_{\mu}+p \,k_{\mu}\big) \quad \Rightarrow \quad A_{\text{M}\,\mu} dx^{\mu}=\frac{Q}{q-p}\big(d\overline{\tau}+p\,q \,d\overline{\sigma})\,.
\end{align}
If we couple this electromagnetic field to each of the metrics in bigravity independently, we obtain what we found in the case of separate matter sectors, where ${{T}_{M}}^{\mu}{}_{\nu} \ne 0\, ,    {\mathcal{T}_{M}}^{\mu}{}_{\nu}\ne 0$, but if we set the electric and magnetic charges equal, what we obtain is ${{T}_{M}}^{\mu}{}_{\nu} = 0=   {\mathcal{T}_{M}}^{\mu}{}_{\nu}$, which also occurs in GR. The same happens if we couple the electromagnetic field to the metrics in bigravity via the effective metric and set $G=Q$ as in (\ref{equal_charges}). This solution satisfied (\ref{gf_evac_eom}). In terms of the effective tensor we have
\begin{align}
    \nonumber \mathcal{L}_{M}&=-\frac{1}{16\pi}{g_{E}{}}^{\mu\rho}{g_{E}{}}^{\nu\sigma}F_{\mu\nu}F_{\rho\sigma}\ne 0\,, \\
    \nonumber {T_{E}{}}_{\mu\nu}& =\frac{1}{4 \pi}\left({g_{E}}^{\rho\sigma}F_{\mu\rho}F_{\nu\sigma}-\frac{1}{4}{g_{E}}_{\mu\nu}{g_{E}{}}^{\gamma\rho}{g_{E}{}}^{\delta\sigma}F_{\gamma\delta}F_{\rho\sigma}\right) = 0\,, \\
    \Rightarrow \quad  {{T}_{M}}^{\mu}{}_{\nu}& =0=  {\mathcal{T}_{M}}^{\mu}{}_{\nu}.
\end{align}

\subsubsection{Proportional metrics}

The charged solution on which we work here is a stationary solution that satisfies the conditions in (\ref{gf_evac_stat_constraints}). We can study this solution using the effective metric. For the case $f_{\mu\nu}=C^{2}g_{\mu\nu}$, both equations in (\ref{gf_evac_stat_constraints}) are equivalent, because the quantities defined with each metric are related, e.g., $\Upsilon^{i}_{f}\propto \Upsilon^{i}_{g}$. These restrictions lead to the equations for this proportional solution as in (\ref{gf_evac_eom}), with each set of equations being equivalent. 

For this solution, we consider the scalar functions in (\ref{scalar_matter_indep}) with the same restrictions for the parameters as in (\ref{constraint_prop}). In order to obtain proportional metrics using the effective metric, we restrict the matter coupling parameters, $\beta \propto \alpha$. Then, the effective metric, $g_{E\,\mu\nu}$ and the electromagnetic tensor are
\begin{align*}
    \beta & = \left(\frac{{\kappa_{g}}^{2}}{{\kappa_{f}}^{2}}\right) C\,\alpha \quad 
 \Rightarrow \quad g_{E\,\mu\nu}= (\alpha+C\beta)^{2}g_{\mu\nu}=\left({\frac{\alpha}{\kappa_{f}^{2}}}\right)^{2}\left(C^{2}\kappa _{g}^{2}+\kappa_{f}^{2}\right)^{2} g_{\mu\nu}   \, , \\  
    \nonumber {T_{E}}_{\mu\nu}&=\frac{1}{\alpha(\alpha+\beta\,C)}{T_{M}}_{\mu\nu}=\frac{\kappa_{f}^{2}}{\alpha^{2}}\frac{1}{\left(C^{2}\kappa _{g}^{2}+\kappa_{f}^{2}\right)}{T_{M}}_{\mu\nu}\, , \quad \mathcal{T}_{M\,\mu\nu} = \left(\frac{\kappa_{g}^{2}}{\kappa_{f}^{2}}\right) {T_{M\,\mu\nu}} \, ,
\end{align*}
with ${T_{E}}_{\mu\nu}$ as in (\ref{em_eff_metric}).  We note that the effective metric depends on the gravitational couplings $\kappa_{g}$ and $\kappa_{f}$, the effective coupling parameter $\alpha$ and on the proportionality constant $C$. In this case we can write down the relation between the charges as
\begin{align}
    \nonumber & Q_{1}=\sqrt{\frac{\alpha}{(C\beta+\alpha)}} Q = \frac{\kappa_{f}}{(C^{2}\kappa_{g}^{2}+\kappa_{f}^{2})^{1/2}} Q= \left(\frac{\kappa_{f}}{\kappa_{g}}\right) Q_{2}=\left(\frac{\kappa_{f}}{\kappa_{g}}\right) \sqrt{\frac{\beta}{C\left(C\beta+\alpha\right)}} Q\, , \\
    \nonumber &G_{1}=\sqrt{\frac{\alpha}{(C\beta+\alpha)}} G=\frac{\kappa_{f}}{(C^{2}\kappa_{g}^{2}+\kappa_{f}^{2})^{1/2}} G= \left(\frac{\kappa_{f}}{\kappa_{g}}\right) G_{2} = \left(\frac{\kappa_{f}}{\kappa_{g}}\right)\sqrt{\frac{\beta}{C\left(C\beta+\alpha\right)}} G \, .
\end{align}
These restrictions between the parameters lead us to
\begin{align}
    \nonumber \mathscr{h}_{\mu\nu}= \left(\frac{\kappa_{g}}{\kappa_{f}}\right) h_{\mu\nu} \, , \, \,  \mathcal{A}_{\mu}= \left(\frac{\kappa_{g}}{\kappa_{f}}\right) A_{\mu} \, , \, \,  \Upsilon_{f}= \left(\frac{\kappa_{g}}{\kappa_{f}}\right) \Upsilon_{g} \,, \, \,  A_{\text{M}\,\mu}=\frac{1}{q^{2}-p^{2}}\big(Q \,q \,l_{\mu}+ G\, p \,k_{\mu}\big)\, ,
\end{align}
where $A_{\text{M}\,\mu}$ is coupled to the metrics via the effective metric. With these relations, we confirm that the equations we obtained for this solution in (\ref{gf_evac_eom}) are equivalent, as the fields, operators and sources are proportional.


\section{Conclusions} \label{FR}

In this work, we presented an analysis of AdS waves and Pleba\'nski-Demia\'nski solutions in bigravity coupled to matter in the framework of the classical Kerr-Schild double copy.  In order to achieve this, we extended the formalism presented in \cite{Monteiro:2014cda,Carrillo-Gonzalez:2017iyj} in GR and single generalized Kerr-Schild ansatz, and apply it to bigravity as in  \cite{Garcia-Compean:2024uie} using a double generalized Kerr-Schild ansatz. We obtained the equations of motion for maximally symmetric backgrounds bigravity implementing a double Kerr-Schild ansatz. In the framework of the classical Kerr-Schild double copy, for bigravity we obtain (\ref{eom_dc_1}) at the level of the double copy, (\ref{eom_sc_1}) at the level of the single copy and (\ref{eom_zc_1}) correspond to the zeroth copy Kerr-Schild classical equations. We studied these equations restricting to specific stationary and time-dependent solutions in bigravity.

The time-dependent solutions we considered are the AdS bigravitational wave solutions from \cite{Ayon-Beato:2018hxz} written in a single Kerr-Schild form, which satisfy the classical double copy equations in bigravity (\ref{time_dep_eom}). For the case of the flat background, the double copy equations represent massless and massive equations for the ``+'' and ``$-$'' spin-2 decoupled fields respectively. In the single copy equations, the ``+'' and ``$-$'' fields satisfy Maxwell and Proca equations respectively for the decoupled fields which are coupled to matter, and for the zeroth copy we obtain a massive and massless Klein-Gordon equation respectively.

The stationary solutions in bigravity we presented are a bigravity analog of the Pleba\'nski-Demia\'nski solutions in GR \cite{plebanski1976rotating} which allow a double Kerr-Schild form when written in Pleba\'nski coordinates. First we presented the double copy of the Pleba\'nski-Demia\'nski solution in GR. Then in bigravity, we use a similar ansatz to the one used in GR for both metrics, given in (\ref{scalar_matter_indep}). The parameters we have for this solution in bigravity are the mass parameter $m_{i}$, NUT parameter $N_{i}$, electric charge $Q_{i}$, magnetic charge $G_{i}$ with $i=1,2$, for each of the metrics. The kinematic parameters $\gamma$ and $\epsilon$ for both metrics are the same and the cosmological constants are related as $\Lambda_{f}=\Lambda_{g}/C^{2}, \Lambda_{g}=\Lambda$, which are the same restrictions used in \cite{Ayon-Beato:2015qtt}. These bigravity solutions presented are written in a double Kerr-Schild form and similar expressions are found to the ones in \cite{Garcia-Compean:2024uie}. 

Pleba\'nski-Demia\'nski solutions satisfy equations (\ref{gf_evac_eom}), which at the level of the double copy are the equations for spin-2 fields in a curved background coupled to matter. The single copy equations are a couple of Maxwell equations coupled to a source and a pair of Klein-Gordon equations coupled to a source, where the inhomogeneous scalar function satisfies (\ref{inhom_eqs}), which is a Klein-Gordon equation with a local coefficient that depends on the spatial coordinates.  Pleba\'nski coordinates reduce the equations in contrast to the global static coordinates used in \cite{Garcia-Compean:2024uie}, where one obtains additional terms which cannot be interpreted in terms of the double copy fields, such terms are also presented in \cite{Bah:2019sda,Alkac:2021bav}. The calculation time in these coordinates was considerably shorter compared  to the one using spheroidal coordinates.

The separate sector permits unrestricted parameters $m_{i}$, $N_{i}$, $Q_{i}$, $G_{i}$ with $i=1,2$, with $\gamma$ and $\epsilon$ equal for both metrics and related cosmological constants, $\Lambda$. These results are related to \cite{Ayon-Beato:2015qtt} for the single Kerr-Schild case, i.e., vanishing NUT and magnetic charge parameters and a change of coordinates. We restrict to the case where the charges are related and proportional metrics. The case where only the charges are related can be interpreted as proportional charges and equal Maxwell fields or equal charges and proportional Maxwell fields. In the case of proportional metrics, the ``$-$'' decoupled field is zero.

For the case of matter fields coupled via the effective metric and stationary solutions, we considered three cases, which consist of electrovacuum for $g_{\mu\nu}$ and vacuum for $f_{\mu\nu}$, and we recover \cite{Garcia-Compean:2024uie,Ayon-Beato:2015qtt} after a change of coordinates. The metric $g_{\mu\nu}$ and the effective metric are proportional and the same relation holds for the electromagnetic tensors constructed with the metrics, so we obtain equivalent results as in the case of separate matter. Moreover, we considered the case where the electric and magnetic charges coincide which result in vanishing electromagnetic tensors, and also the case of proportional metrics. A case with non-trivial electrovacuum for both metrics was not discussed in the present article in terms of the effective metric.

In the separate matter and when the matter is coupled using the effective metric, in the case of proportional metrics we obtain proportional equations of motion and that ``$-$'' fields in (\ref{redef_ads_2}) vanish at the level of the double, single and zeroth copy. All the solutions we studied here are within the formalism of the double Kerr-Schild in bigravity coupled to matter, and we can reduce to the single Kerr-Schild ansatz.

For further work, the study of the sourced Weyl double copy can provide tools to better understand the results presented in this paper in the context of the double copy, even in Einstein-Maxwell theory, as was studied in \cite{Armstrong-Williams:2024bog}, and then consider massive gravity solutions. We would like to address this problem in the near future.

With this work we extended the framework of the classical Kerr-Schild double copy by considering double Kerr-Schild solutions coupled to matter, namely the Pleba\'nski-Demia\'nski solutions, widening the application of the KS double copy relations in massive gravity.

\acknowledgments
It is a pleasure to thank E. Ayón-Beato and  A. Luna for useful feedback and enlightening comments. C. Ramos thanks Conahcyt for the scholarship No. 833288.




\clearpage




\end{document}